\documentclass[12pt,reqno]{article}
\usepackage{amsmath}
\usepackage{amsfonts}
\usepackage{hyperref}

\hypersetup{colorlinks,linkcolor=blue,citecolor=red,urlcolor=blue,%
bookmarksopen,pdftitle={Fivebrane Instanton Corrections to the
Universal Hypermultiplet},pdfauthor={Marijn Davidse, Ulrich Theis,
Stefan Vandoren}}

\allowdisplaybreaks
\numberwithin{equation}{section}

\renewcommand{\d}{\mathrm{d}}
\newcommand{\I}{\mathrm{i}}
\newcommand{\e}{\mathrm{e}}
\newcommand{\p}{\partial}

\newcommand{\cD}{\mathcal{D}}
\newcommand{\cE}{\mathcal{E}}
\newcommand{\cF}{\mathcal{F}}
\newcommand{\cG}{\mathcal{G}}
\newcommand{\cH}{\mathcal{H}}
\newcommand{\cL}{\mathcal{L}}
\newcommand{\cM}{\mathcal{M}}

\newcommand{\ab}{{\bar{a}}}
\newcommand{\bb}{{\bar{b}}}

\newcommand{\ga}{\gamma}
\newcommand{\de}{\delta}
\newcommand{\ep}{\varepsilon}
\newcommand{\eps}{\epsilon}
\newcommand{\beps}{\bar{\epsilon}}
\newcommand{\si}{\sigma}
\newcommand{\bsi}{\bar{\sigma}}
\newcommand{\la}{\lambda}
\newcommand{\bla}{\bar{\lambda}}
\newcommand{\bxi}{\bar{\xi}}
\newcommand{\bpsi}{\bar{\psi}}

\newcommand{\SU}{\mathrm{SU}}
\newcommand{\SO}{\mathrm{SO}}
\newcommand{\SL}{\mathrm{SL}}
\newcommand{\Oo}{\mathrm{O}}
\newcommand{\U}{\mathrm{U}}

\newcommand{\2}{\sqrt{2\,}}
\newcommand{\half}{\tfrac{1}{2}}
\newcommand{\quart}{\tfrac{1}{4}}

\newcommand{\+}{\mskip1mu}
\newcommand{\tab}{\quad\,}
\newcommand{\lp}{\begin{pmatrix}}
\newcommand{\rp}{\end{pmatrix}}

\newcommand{\W}[2]{W_{\!#1}^{#2}}
\newcommand{\bW}[2]{\bar{W}_{\!#1}^{#2}}
\newcommand{\G}[3]{\Gamma_{\!#1}{}^{#2}{}_{#3}}
\providecommand{\slsh}[1]{#1\!\!\!\!/\,}

\newcommand{\cl}{\mathrm{cl}}
\newcommand{\inst}{\mathrm{inst}}
\newcommand{\eff}{\mathrm{eff}}

\newcommand{\arxiv}[1]{\href{http://arxiv.org/abs/#1}{\texttt{#1}}}

\DeclareMathOperator{\Tr}{Tr}
\DeclareMathOperator{\Ind}{Ind}

\DeclareSymbolFont{AMSb}{U}{msb}{m}{n}
\DeclareMathSymbol{\fieldC}{\mathalpha}{AMSb}{"43}
\DeclareMathSymbol{\fieldR}{\mathalpha}{AMSb}{"52}
\DeclareMathSymbol{\fieldZ}{\mathalpha}{AMSb}{"5A}

\begin{document} 

\setcounter{page}{0}
\thispagestyle{empty}

\begin{flushright} \small
 ITP--UU--04/12 \\ SPIN--04/06 \\ TUW--04--09 \\
 \href{http://arxiv.org/abs/hep-th/0404147}{hep-th/0404147}
\end{flushright}
\smallskip

\begin{center} \LARGE
 Fivebrane Instanton Corrections \\ to the \\ Universal Hypermultiplet
 \\[2ex] \normalsize
 Marijn Davidse$^1$, Ulrich Theis$^2$ and Stefan Vandoren$^1$ \\[2ex]
 {\small\slshape
 $^1$Institute for Theoretical Physics \emph{and} Spinoza Institute \\
 Utrecht University, 3508 TD Utrecht, The Netherlands \\
 \href{mailto:M.Davidse@phys.uu.nl}{\upshape\ttfamily M.Davidse,}
 \href{mailto:S.Vandoren@phys.uu.nl}{\upshape\ttfamily
 S.Vandoren@phys.uu.nl} \\[1ex]
 $^2$Institute for Theoretical Physics, Vienna University of
 Technology, \\ Wiedner Hauptstrasse 8--10, A-1040 Vienna, Austria \\
 \href{mailto:theis@hep.itp.tuwien.ac.at}{\upshape\ttfamily
 theis@hep.itp.tuwien.ac.at}}
\end{center}
\vspace{6ex}

\hrule\bigskip

\centerline{\bfseries Abstract} \medskip

We analyze the Neveu-Schwarz fivebrane instanton in type IIA string
theory compactifications on rigid Calabi-Yau threefolds, in the
low-energy supergravity approximation. It there appears as a finite
action solution to the Euclidean equations of motion of a double-tensor
multiplet (dual to the universal hypermultiplet) coupled to $N=2$, $D=4$
supergravity. We determine the bosonic and fermionic zero modes, and the
single-centered instanton measure on the moduli space of collective
coordinates. The results are then used to compute, in the semiclassical
approximation, correlation functions that nonperturbatively correct the
universal hypermultiplet moduli space geometry of the low-energy
effective action. We find that only the Ramond-Ramond sector receives
corrections, and we discuss the breaking of isometries due to
instantons.

\bigskip\hrule

\newpage

\small
\tableofcontents{}
\normalsize

\section{Introduction}

One particular object that plays an important role in nonperturbative
string theory is the Neveu-Schwarz fivebrane (NS5-brane). It is the
magnetic dual of the fundamental string, and appears as a soliton-like
solution preserving half of the supersymmetry, for instance in heterotic
\cite{S-CHS1-DL} and type II strings \cite{CHS2}. Compared to the type
IIB theory, the NS5-brane in IIA is particularly intriguing because,
unlike D-branes, strings cannot end on it. This makes it difficult to
construct the worldvolume theory, or to compute the partition function
of the fivebrane, see e.g.\ \cite{DVV} and references therein.

The transverse space to the fivebrane is four-dimensional with
coordinates $x^\mu$, $\mu=1,\dots,4$, and as a solution to the
supergravity equations of motion, it is characterized by the dilaton
and the NS 2-form field strength,
 \begin{equation} \label{NS5-brane}
  \e^{2\phi} = \e^{2\phi_\infty} + \frac{Q}{r^2}\ ,\qquad H_{\mu\nu\rho}
  = \ep_{\mu\nu\rho}{}^\si\, \nabla_{\!\si} \phi\ .
 \end{equation}
Here $r^2=|x|^2$, and $Q$ is the fivebrane charge, which is quantized
in appropriate units. The four-dimensional spacetime metric is
(conformally) flat, and we have set the gauge fields, present in the
heterotic string, or any other fields from the Ramond-Ramond sector in
type II, to zero. See \cite{S-CHS1-DL,CHS2} for more details.

Clearly, the solution \eqref{NS5-brane} is localized in the transverse
four-dimensional Euclidean space, with the center chosen to be at the
origin. It is easy to generalize \eqref{NS5-brane} to multicentered
solutions that are localized at a finite number of points in
$\fieldR^4$. From a four-dimensional point of view, one expects
therefore the NS5-brane to be a (multi-centered) instanton. The
remaining six dimensions, in the direction of the fivebrane, can be
chosen non-compact and flat, but may also be replaced by a compact
$T^6$, $K3\times T^2$, or a Calabi-Yau (CY) threefold, around which the
(Euclidean) fivebrane is wrapped. For theories with more than $N=2$
supersymmetry in four dimensions, such instanton effects contribute to
higher derivative terms in the effective action, as was shown for
heterotic strings on $T^6$ in \cite{HM-GKKOPP}, or for type II
compactifications on $K3\times T^2$ in \cite{KOP}. For a nice review,
see \cite{Ki}. The case of $N=2$, corresponding to type II strings on
CY-threefolds, is most interesting, as now the fivebrane instantons
contribute to the low-energy effective action \cite{BBS}. The aim of
this paper is to compute some of these instanton effects explicitly in
the semiclassical and supergravity approximation. A similar study for
heterotic strings on CY-threefolds, with $N=1$ supersymmetry in four
dimensions, was initiated in \cite{R}. Computations of nonperturbative
superpotentials due to fivebrane instantons were given in \cite{W}.

In this paper, we thus focus on Neveu-Schwarz fivebrane instantons in
type IIA string theory compactified on a CY manifold. In the absence of
internal fluxes, a generic such compactification with Hodge numbers
$h_{1,1}$ and $h_{1,2}$ has as its low-energy effective action
four-dimensional $N=2$ supergravity coupled to $h_{1,1}$ vector
multiplets, $h_{1,2}$ hypermultiplets, and one tensor multiplet that
contains the dilaton \cite{BCF}. The simplest situation is therefore
to freeze all the vector multiplets (they play no significant role in
the present discussion since the other matter multiplets are neutral
in the absence of fluxes) and to set $h_{1,2}=0$. For a more general
discussion on the geometry of vector and hypermultiplet moduli spaces,
we refer to the review \cite{A} and references therein.

Calabi-Yau manifolds with no complex structure moduli ($h_{1,2}=0$) are
called rigid. Truncating the vector multiplets, the resulting
four-dimensional low-energy effective action is that of a tensor
multiplet coupled to $N=2$ supergravity, and the bosonic part of the
Lagrangian at string tree-level is given by\footnote{Throughout this
paper, we work in units in which Newton's constant $\kappa^{-2}=2$,
except in section \ref{sect_rigid}, where we compare rigid and local
$N=2$ scalar-tensor couplings.}
 \begin{align} \label{TM-action}
  e^{-1} \cL_\mathrm{T} & = - R - \frac{1}{2}\, \p^\mu \phi\, \p_\mu
    \phi + \frac{1}{2}\, \e^{2\phi} H^\mu H_\mu \notag \\*
  & \tab - \frac{1}{4}\, F^{\mu\nu} F_{\mu\nu} - \frac{1}{2}\,
    \e^{-\phi} \big( \p^\mu \chi\, \p_\mu \chi + \p^\mu \varphi\,
    \p_\mu \varphi \big) - \frac{1}{2}\, H^\mu \big( \chi \p_\mu \varphi
    - \varphi \p_\mu \chi \big)\ ,
 \end{align}
where $H^\mu=\tfrac{1}{6}\,\ep^{\mu\nu\rho\si}H_{\nu\rho\si}$ is the
dual NS 2-form field strength. The first line comes from the NS sector
in ten dimensions, and $\phi$ together with $H^\mu$ forms an $N=1$
tensor multiplet that is also present in heterotic compactifications.
It is then not surprising that the NS5-brane solution \eqref{NS5-brane}
straightforwardly descends to a finite action solution of the Euclidean
tensor multiplet Lagrangian, with RR fields set to zero \cite{R,BBS,BB}.
The second line descends from the RR sector. In particular, the
graviphoton with field strength $F_{\mu\nu}$ descends from the
ten-dimensional RR 1-form, and $\varphi$ and $\chi$ can be combined
into a complex scalar $C$ that descends from the holomorphic components
of the RR 3-form with (complex) indices along the holomorphic 3-form of
the CY. Notice the presence of constant shift symmetries on both $\chi$
and $\varphi$. Together with a rotation on $\chi$ and $\varphi $ they
form a three-dimensional subgroup of symmetries. The presence of the RR
scalars then opens up the possibility to construct more general
instanton solutions for which the RR sector becomes nontrivial. We will
extensively discuss what kind of RR backgrounds the NS5-brane instanton
can support, and how the correlation functions depend on it. We should
stress that these instantons are distinct from membrane instantons,
arising from wrapping Euclidean D2-branes around three-cycles in the CY
\cite{BBS}. Membrane instantons have a different dependence on the
string coupling constant, and have different charges. Their study is
beyond the scope of the present paper.

The tensor multiplet Lagrangian \eqref{TM-action} is dual to the
universal hypermultiplet. This can be seen by dualizing the 2-form
into an axionic pseudoscalar field $\sigma$, after which one obtains
(modulo a surface term)
 \begin{align}\label{UH-action}
  e^{-1} \cL_\mathrm{UH} & = - R - \frac{1}{4}\, F^{\mu\nu} F_{\mu\nu}
    - \frac{1}{2}\, \p^\mu \phi\, \p_\mu \phi - \frac{1}{2}\,
    \e^{-\phi} \big( \p^\mu \chi\, \p_\mu \chi + \p^\mu \varphi\,
    \p_\mu \varphi \big) \notag \\*
  & \tab - \frac{1}{2}\, \e^{-2\phi} \big( \p_\mu \si + \chi \p_\mu
    \varphi \big)^2 \ .
 \end{align}
The four scalars define the classical universal hypermultiplet at
string tree-level, a non-linear sigma model with a quaternion-K\"ahler
target space $\SU(1,2)/\U(2)$ \cite{CFG}. This target space has an
$\SU(1,2)$ group of isometries, with a three-dimensional Heisenberg
subalgebra that generates the following shifts on the fields,
 \begin{equation} \label{Heis-alg}
  \phi \rightarrow \phi \ ,\qquad \chi \rightarrow \chi + \gamma\ ,
  \qquad \varphi \rightarrow \varphi + \beta\ ,\qquad \sigma
  \rightarrow \sigma - \alpha - \gamma\, \varphi\ ,
 \end{equation}
where $\alpha$, $\beta$, $\gamma$ are real (finite) parameters. A fourth
symmetry is again a rotation on $\varphi$ and $\chi$, which is now
accompanied by a compensating transformation on $\sigma$. The remaining
four isometries involve non-trivial transformations on the dilaton, and
hence will change the string coupling constant.

Quantum corrections, both perturbative and nonperturbative, will break
some of the isometries and alter the classical moduli space of the
universal hypermultiplet, while keeping the quaternion-K\"ahler
property intact, as required by supersymmetry \cite{BW}. At the
perturbative level, the authors of \cite{AMTV} reconsidered the
analysis of \cite{S,GHL}, and found a non-trivial one-loop correction
that modifies the low-energy tensor multiplet Lagrangian
\eqref{TM-action}, or after dualization, the universal hypermultiplet
Lagrangian \eqref{UH-action}. More recently, this one-loop correction
was written and analyzed in the language of projective superspace in
\cite{ARV}, using the tools developed in \cite{dWRV}. At the
nonperturbative level, not much is known explicitly at present. Some
earlier references on this topic are \cite{BBS,GMV,OV,BB,GS,K}. The
problem of finding the quantum corrections to the hypermultiplet moduli
space metric is somewhat similar to determining the hyperk\"ahler metric
on the Coulomb branch of three-dimensional gauge theories with eight
supercharges \cite{SW,DKMTV}. This analogy was used in \cite{ARV} to
conjecture a natural candidate metric on the universal hypermultiplet
moduli space induced by fivebrane instantons. Given the metric, one can
then investigate the properties of the scalar potential that arises
after gauging the remaining unbroken isometries. An interesting
application, based on a proposal for the membrane instanton corrections
to the universal hypermultiplet moduli space metric \cite{K}, was
recently given in the context of finding de Sitter vacua from $N=2$
gauged supergravity \cite{BM}.

In this paper, elaborating on our earlier work in \cite{TV1,DDVTV}, we
take the first steps in computing the universal hypermultiplet moduli
space metric explicitly by using semiclassical instanton calculations.
As is familiar from Seiberg-Witten theory, and its three-dimensional
version \cite{SW,DKMTV}, one has to be careful in matching the
coordinates on the moduli space with the fields from the supergravity
(or string) theory. \label{sec:hier} In fact, using semiclassical
techniques, we can only determine this relation at weak coupling, and
one can think of the asymptotic value of the dilaton (the string
coupling constant) as a radial coordinate on the moduli space. Away from
the semiclassical regime, or equivalently, the asymptotic region of the
moduli space, the relation between the fields and coordinates is
expected to be more complicated, and using our approach, we have no
access to this regime. Therefore, in this paper, we have to content
ourselves with computing only the first non-vanishing, but leading
exponential correction in the full moduli space metric. Subleading
corrections have to be computed using other methods, or can perhaps be
fixed by the (super) symmetry constraints (like for instance the
quaternionic geometry) and the regularity assumptions of the moduli
space.

This paper is organized as follows: In section \ref{dtm-section} we
introduce the double-tensor multiplet dual to the universal
hypermultiplet. Section \ref{Fivebrane-inst} contains a review of the
fivebrane instanton and anti-instanton solutions present in the
double-tensor multiplet, and we discuss their properties. Section
\ref{sect_rigid} deals with the generic form of the effective action for
scalar-tensor interactions, and we discuss the Euclidean supersymmetry
rules which we apply to the universal double-tensor multiplet in section
\ref{sect_DTM}. In section \ref{sect_unbroken}, we show explicitly that
the fivebrane instanton preserves half of the (Euclidean)
supersymmetry, and we use the broken supersymmetries to determine the
fermionic zero modes in section \ref{sect_FZM}, both for instantons and
anti-instantons. A crucial ingredient of any instanton calculation is
finding the instanton measure on the space of bosonic and fermionic
collective coordinates. For single-centered instantons, we determine
this measure in section \ref{sect_measure}. The one-loop determinant is
left unspecified, calculating it is beyond the scope of this paper. All
this is preparatory material, and in section \ref{sect_correl} we
finally compute the instanton induced correlation functions, from which
we determine the asymptotic corrections to the universal hypermultiplet
moduli space in section \ref{sect_modul}. Not surprisingly, since the
metric and other quantities for hypermultiplets are not holomorphic,
contributions are found from both instantons and anti-instantons. We
further discuss the structure of the isometry group, and the breaking
of isometries to discrete subgroups, due to fivebrane instantons.

We end this paper with some conclusions and remarks for further
investigation. To make this paper as readable as possible, we have
included several appendices with additional information, conventions
and technical details.

\section{The universal double-tensor multiplet} \label{dtm-section}

Instead of dualizing the tensor multiplet Lagrangian \eqref{TM-action}
to a hypermultiplet, we can also use the shift symmetry of one of the RR
scalars, say $\varphi$, to dualize to a double-tensor multiplet. This
can only be done if the shift symmetry survives in the full quantum
theory. In the presence of only NS5-brane instantons, this is indeed the
case. We will come back to the discussion of broken isometries below.
After dualization\footnote{We are dualizing here at the classical level.
Possible quantum corrections coming from the path integral measure are
not taken into account.}, the resulting tree level double-tensor
multiplet Lagrangian reads \cite{TV1,TV2}
 \begin{equation} \label{DTM-action}
  e^{-1} \cL_\mathrm{DT} = - R - \frac{1}{4}\, F_{\mu\nu} F^{\mu\nu}
  - \frac{1}{2}\, \p^\mu \phi\, \p_\mu \phi - \frac{1}{2}\, \e^{-\phi}\,
  \p^\mu \chi\, \p_\mu \chi + \frac{1}{2}\, M^{IJ} H^\mu_I H_{\mu J}\ ,
 \end{equation}
where the $H_I$ are a pair of 3-form field strengths, $H^\mu_I=\half
\ep^{\mu\nu\rho\si}\p_\nu B_{\rho\si I}$, and
 \begin{equation}
  M^{IJ} = \e^{\phi} \begin{pmatrix} 1 & - \chi \\[2pt] - \chi &
  \e^{\phi} + \chi^2 \end{pmatrix}\ .
 \end{equation}
The two scalars $\phi$ and $\chi$ parameterize the coset $\SL(2,\fieldR)
/\Oo(2)$. More on the geometry of the target space before and after
dualization can be found in appendix \ref{cosets}. The presence of the
tensors breaks the SL$(2,\fieldR)$ symmetries to a two-dimensional
subgroup generated by rescalings of the tensors and of $\tau\equiv\chi
+2\I\,\e^{\phi/2}$, and by the generator that acts as a shift on $\chi$
and transforms the tensors linearly into each other \cite{TV1,DDVTV},
 \begin{equation} \label{shift-chi}
  \chi \rightarrow \chi + \gamma\ ,\qquad B_1 \rightarrow B_1 + \gamma
  B_2\ ,
 \end{equation}
with $\phi$ and $B_2$ invariant. This symmetry played a crucial role
in determining the perturbative one-loop correction to the universal
hypermultiplet. We will not discuss this perturbative correction
any further in this paper, but refer to \cite{AMTV,ARV}.

In \cite{TV1,DDVTV}, Bogomol'nyi equations were derived and solved for
the double-tensor multiplet action (\ref{DTM-action}). The solutions
were shown to describe fivebrane and membrane-like instantons, and in
this paper we focus on the fivebrane solutions. The semiclassical and
supergravity description of these instantons is most natural in the
Euclidean continuation of the double-tensor multiplet action. As
explained in \cite{TV1}, this has the advantage over the Euclidean
universal hypermultiplet or tensor multiplet in that the Euclidean
action is semi-positive definite. Hence, a saddle-point approximation
can be justified and a Bogomol'nyi bound can be derived. Both the
tensor- and hypermultiplet actions contain pseudoscalars which have
negative kinetic energy after analytic continuation. In such a
formulation, the semiclassical approximation is hard to justify, and
one has to add surface terms that guarantee the stability of the
Euclidean hypermultiplet action. Dualizing the pseudoscalars into
tensors leads to a formulation with a manifestly positive semi-definite
action, and precisely produces the surface terms needed to make the
pseudoscalar action bounded. This subtlety is similar to the
D-instanton in type IIB supergravity in ten dimensions, where the axion
is best dualized to a 9-form field strength \cite{GGP}. See also
\cite{CLLPST} for related issues. For these reasons, we set up our
instanton calculation in the double-tensor multiplet Lagrangian.

Perturbatively, the double-tensor multiplet guarantees $\U(1)\times
\U(1)$ isometries in the dual hypermultiplet description.
Nonperturbatively, however, the duality also involves the constant
modes of the dual scalars $\varphi$ and $\sigma$ by means of
theta-angle-like terms. These are surface terms which have to be added
to the double-tensor multiplet Lagrangian, and which are non-vanishing
in the presence of instantons and anti-instantons. In Euclidean space,
they can be written as integrals over 3-spheres at infinity,
 \begin{equation} \label{theta-angles}
  S^E_\mathrm{surf} = -\I \Big( \varphi\! \int_{S^3_\infty}\!\!\! H_1
  + \sigma\! \int_{S^3_\infty}\!\!\! H_2 \Big) = -\I\+ \varphi\+ Q_1
  - \I\+ \si\+ Q_2\ .
 \end{equation}
Here, $Q_1$ and $Q_2$ are related to the membrane- (which we do not
discuss in this paper) and fivebrane-instanton charges associated with
the two tensors $H_1$ and $H_2$, and $\varphi$ and $\sigma$ are some
parameters. In the dual hypermultiplet theory, they play the role of
coordinates on the moduli space. The dualization is done by promoting
$\sigma$ and $\varphi$ to fields that serve as Lagrange multipliers
enforcing the Bianchi identities on the tensors. Boundary terms such as
\eqref{theta-angles} are also added in three-dimensional gauge theories
in the Coulomb phase \cite{P,SW,DKMTV}. There, the effective theory can
be described in terms of a vector that can be dualized into a scalar
(the dual photon) along the same lines as described above. Clearly,
depending on which charges are turned on (membrane or fivebrane), the
shift symmetry in $\varphi$ or $\sigma$ (or some linear combination
thereof) will be broken to a discrete subgroup. We will discuss the
breaking of isometries in detail in section \ref{sect_modul}.

\section{Fivebrane instantons} \label{Fivebrane-inst}

Our strategy is to perform a semiclassical instanton calculation in the
double-tensor multiplet formulation, in the supergravity approximation.
We should therefore first discuss the properties of the NS5-brane
instanton, as a solution of the Euclidean equations of motion of the
Lagrangian \eqref{DTM-action}. As mentioned in the previous section,
the equations that determine the instanton solutions were found in
\cite{TV1,DDVTV} by deriving a Bogomol'nyi bound. The fivebrane
instanton solution satisfies the Bogomol'nyi equation \cite{TV1}
 \begin{equation} \label{O2BPS}
  \lp H_{\mu1} \\[2pt] H_{\mu2} \rp = \pm\, \p_\mu \lp \e^{-\phi} \chi
  \\[2pt] \e^{-\phi} \rp\ ,
 \end{equation}
where the plus and minus signs refer to instantons or anti-instantons,
respectively. It will often be useful to change basis and define
$\hat{H}_1=H_1-\chi H_2$, which satisfies the Bogomol'nyi equation
$\hat{H}_{\mu1}=\pm\e^{-\phi}\,\p_\mu\chi$. Such field configurations
have vanishing energy-momentum tensors, so they must live in Ricci-flat
spaces. In this paper we shall consider only flat space with metric
$g_{\mu\nu}=\delta_{\mu\nu}$. Furthermore, we choose a vanishing
graviphoton $F_{\mu\nu}$, and therefore we only focus on the
scalar-tensor sector. The second equation in \eqref{O2BPS} comes from
the NS sector and specifies the NS5-brane instanton. The first equation
determines the RR sector and characterizes the RR background in which
the fivebrane instanton lives. We now proceed to solve these equations.

The closure of the 3-form field strengths implies Laplace-like
equations for the scalars. There are two approaches to proceed and solve
these equations, which should be equivalent. The first one is to find
solutions on the whole $\fieldR^4$. This can only be done if appropriate
source terms are added to the equations of motion. The other approach,
which we will follow here, is to excise points $\{x_i\}$ from
$\fieldR^4$, the locations of the instantons. On such a space, we find
the multi-centered solutions in terms of two harmonic functions,
 \begin{equation} \label{5-inst-sol}
  \e^{-\phi} = \e^{-\phi_\infty} + \sum_i\, \frac{|Q_{2i}|}{4\pi^2\,
  (x-x_i)^2}\ ,\qquad \e^{-\phi} \chi = \e^{-\phi_\infty} \chi_{\infty}
  + \sum_i\, \frac{Q_{1i}}{4\pi^2\, (x-x_i)^2}\ ,
 \end{equation}
where $Q_{1i},Q_{2i}$, $\chi_\infty$, and $\phi_\infty$ are integration
constants; the latter two determine the asymptotic values of the fields
at infinity. The absolute values of $Q_{2i}$ appear to make $\e^{-\phi}$
positive everywhere in space. We identify the string coupling constant
via $g_s=\e^{-\phi_\infty/2}$. Furthermore, two charges are defined by
integrating the tensor field strengths $H_{\mu\nu\rho\,I}=-\ep_{\mu\nu
\rho\si}H^\si_I$ over 3-spheres at infinity,
 \begin{equation}
  Q_I = \int_{S^3_\infty}\! H_I\ ,\quad I=1,2\ .
 \end{equation}
They are related to the constants appearing in the scalar fields through
the field equation \eqref{O2BPS}. Using $**=-1$ on a 3-form in four
Euclidean dimensions, we find
 \begin{equation}
  Q_2 = \mp \sum_i |Q_{2i}|\ ,\qquad Q_1 = \mp \sum_i Q_{1i}\ .
 \end{equation}
This implies that for instantons, $Q_2$ should be taken negative,
whereas for anti-instantons, $Q_2$ must be positive\footnote{Charges
$Q'_{Ii}$ can also be defined as the integrals of the field strengths
over a 3-sphere around $x_i$. For anti-instantons they coincide with
$Q_{1i}$ and $|Q_{2i}|$, but for instantons, there are minus signs. To
avoid heavy notation, we will never write $Q'_{Ii}$.}. Note that there
is no restriction on the signs of the $Q_{1i}$.

The (anti-) instanton action for the fivebrane was found to be the
following integral over the boundary of Euclidean spacetime \cite{TV1}
 \begin{equation}
  S_\cl = \pm \int_{\p\cM}\, \big[ \chi H_1 - (\e^{\phi} + \half
  \chi^2)\, H_2 \big]\ ,
 \end{equation}
where the boundary $\p\cM$ consists of the disjoint union of a sphere
at infinity and infinitesimal spheres around the excised points $x_i$.
The value of the action was computed and interpreted for general
multicentered solutions in \cite{DDVTV}. It is finite only if the
values of $\chi$ at the excised points $x_i$ are finite. We have
 \begin{equation}
  \chi_i \equiv \lim_{x\rightarrow x_i} \chi(x) = \frac{Q_{1i}}
  {|Q_{2i}|}\ ,
 \end{equation}
which is finite whenever $Q_{2i}\neq 0$ for nonvanishing $Q_{1i}$.
The finiteness condition can be rewritten as the vanishing of the
charges
 \begin{equation} \label{nomembrane}
  \hat{Q}_{1i} \equiv Q_{1i} - \chi_i\, |Q_{2i}| = 0\ .
 \end{equation}
Plugging the solution into the action, we find
 \begin{equation} \label{inst-act}
  S_\cl = \frac{|Q_2|}{g_s^2} + \frac{1}{2} \sum_i\, |Q_{2i}|\,
  (\chi_\infty - \chi_i)^2\ .
 \end{equation}
The $1/g_s^2$ dependence is consistent with the string theory
expectations for a wrapped NS fivebrane around the entire CY \cite{BBS}.

For a single-centered solution, finiteness of the action requires
 \begin{equation} \label{hatQ1}
  \hat{Q}_1 \equiv Q_1 - \chi_0 Q_2 = 0\ ,
 \end{equation}
for both instantons and anti-instantons. Here, $\chi_0$ is the value of
the RR scalar at the location $x_0$ of the (anti-) instanton. Notice
that, in contrast to the dilaton, the RR scalar remains finite at the
origin. This regularity implies that the field equation for $\chi$
does not need a source. One should therefore think of $\chi_0$ as some
constant RR background flux in which the NS5-brane instanton with
charge $Q_2$ lives. By means of \eqref{nomembrane}, the
quantization of the charges implies a quantization condition on the
value of $\chi_0$. In \cite{DDVTV}, $\hat Q_{1}$ was interpreted as a
membrane instanton charge, because the corresponding instanton action
is linear (instead of quadratic) in the inverse string coupling
constant, in accordance with \cite{BBS}. Therefore our solutions
correspond to purely fivebrane instantons, with the addition of some
RR flux $\chi_0$, but with zero membrane charge. Notice that the
membrane charge $\hat Q_{1}$ is invariant under the symmetry
\eqref{shift-chi}. The action for the single-centered fivebrane
instanton then becomes
 \begin{equation} \label{one-inst-act}
  S_\cl = |Q_2|\, \Big( \frac{1}{g_s^2} + \frac{1}{2}\, (\Delta\chi)^2
  \Big)\ ,
 \end{equation}
where $\Delta\chi=\chi_\infty-\chi_0$. The second term is the
consequence of turning on a nontrivial RR background. This has the net
effect of raising the instanton action by an amount of $(\Delta\chi)^2$.
It vanishes when the boundary values of $\chi$ at $r=0$ and $r=\infty$
are equal, i.e.\ when $\chi_\infty=\chi_0$. Plugging this into the
solution \eqref{5-inst-sol}, one easily checks that $\chi$ is constant
everywhere, $\chi=\chi_\infty$. This is the most trivial RR background,
and the easiest in which we can do instanton calculations. We will
however also consider nonconstant backgrounds, and study how the
fivebrane instanton responds to it.

Notice that the value of the action is positive and the same for
instantons and anti-instantons. To distinguish instantons from
anti-instantons, we add the theta-angle-like terms discussed in the
introduction in \eqref{theta-angles}. Using \eqref{hatQ1}, we can
rewrite this as
 \begin{equation} \label{theta-angles2}
  S_\mathrm{surf} = -\I\+ \sigma\+ Q_2 - \I\+ \varphi\+ Q_1 = \pm\, \I
  (\sigma + \chi_0\, \varphi)\, |Q_2| - \I\+ \varphi\+ \hat{Q}_1\ .
 \end{equation}
The total instanton action is then
 \begin{equation}
  S_\inst^\pm = S_\cl + S_\mathrm{surf}\ ,
 \end{equation}
and will contribute to any instanton-induced correlation function by
exponentiation. After dualization to the universal hypermultiplet,
this instanton action will contribute to the effective moduli space
metric, for which $\phi$, $\chi$, $\varphi$ and $\sigma$ can be thought
of as the coordinates. As a consequence, some of the isometries will
be broken, as we discuss in section \ref{sect_modul}.

\medskip

In section \ref{sect_unbroken} we show that, although these solutions
satisfy the Bogomol'nyi bound, not all of them preserve half of the
supersymmetries. For that, there will be further restrictions on the
multi-centered instanton solutions, namely that all the $\chi_i$ are
equal. This implies that the solution is characterized in terms of a
single harmonic function. When all the $\chi_i$ are equal,
\eqref{inst-act} reduces to the action of a single-centered instanton.
As argued in \cite{DDVTV}, the generic multi-centered solution is
expected to be metastable and decays into a solution where all $\chi_i$
are equal. The value of the action is then lowered to
\eqref{one-inst-act}, and the points $x^i$ can be brought together to
obtain a spherically symmetric solution, without further changing the
value of the action.

The actual instanton calculation we perform in later sections of this
paper will be based on the single-centered instanton. The multi-centered
ones can be written as a multipole expansion, in which the dominant term
contributing to the low-energy effective action is the single-centered
one. Before we compute these instanton effects, we must first discuss
the general form of low-energy effective actions, and the (Euclidean)
supersymmetry transformation rules. The latter are needed to discuss the
BPS properties of our instanton solutions.

\section{Low energy effective actions for scalar-tensor multiplets}
\label{sect_rigid}

The $N=2$ scalar-tensor system consists of $n_T$ tensors $B_{\mu\nu I}$,
$I=1,\dots,n_T$ and $4n-n_T$ scalars $\phi^A$, together with $2n$
two-component spinors $\la^a$. Its self-interactions and its couplings
to $N=2$ supergravity in spacetimes with Lorentz signature have been
derived in \cite{TV2}. We will not repeat it here, but only discuss its
analytic continuation to Euclidean signature. The case of interest for
our applications corresponds to $n=1$ and $n_T=2$, but the discussion
below is for arbitrary scalar-tensor systems. The Wick rotation can
easily be done by using the prescription that under $x^0\rightarrow-\I
x^4$ we have
 \begin{equation} \label{WickB}
  B_{0i} \rightarrow \I B_{4i}\ ,\qquad B_{ij} \rightarrow B_{ij}\ ,
  \qquad i=1,2,3\ .
 \end{equation}
More conventions on spinors and the Euclidean Clifford algebra are
given in appendix \ref{conventions}.

\bigskip
\textbf{Rigid Supersymmetry} 
\medskip

The Euclidean action can be found by doing the Wick rotation as
described above. After extracting an overall minus sign, such that
the action enters the path integral via $\e^{-S}$, it is given by
\begin{align}\label{Eucl-rigid-action}
  \cL & = \frac{1}{2}\, M^{IJ} H^\mu_I H_{\mu J} + \frac{1}{2}\,
    \cG_{AB}\, \p^\mu \phi^A\, \p_\mu \phi^B - \I A_A^I\, H^\mu_I\,
    \p_\mu \phi^A + \frac{\I}{2}\, h_{a\ab}\, \big( \la^a \si^\mu
    \overset{\leftrightarrow}{\cD}_\mu \bla^\ab \big) \notag \\*
  & \tab + \I H_{\mu I} M^{IJ} k_{Ja\ab}\, \la^a \si^\mu \bla^\ab -
    \frac{1}{4}\, V_{ab\,\ab\bb}\, \la^a \la^b\, \bla^\ab
    \bla^\bb\ ,
 \end{align}
for some functions $M^{IJ}(\phi)$, $\cG_{AB}(\phi)$, etc. We denote
$H^\mu_I=\half\ep^{\mu\nu\rho\si}\p_\nu B_{\rho\si I}$, and the
covariant derivative is given by
 \begin{equation} \label{cov-der}
  \cD_\mu \la^a = \p_\mu \la^a + \p_\mu \phi^A\, \G{A}{a}{b}\, \la^b\ .
 \end{equation}
The connection ensures covariance with respect to fermion frame
reparametrizations $\la^a\rightarrow S^a{}_b(\phi)\la^b$.

In the above we have assumed that all the fields $\phi^A$ stay inert
under the Wick rotation. This guarantees that, for a positive definite
metric $\cG_{AB}$, the kinetic terms for the scalars contribute
positively to the action. Sometimes, when dealing with pseudoscalars,
one could pick up a factor of $\I$ upon the Wick rotation such that the
action contains negative kinetic terms. We assume that all such fields
with negative kinetic energy can be dualized into tensors, and hence we
have a positive kinetic energy for the tensors, with positive definite
$M^{IJ}$ (this assumption is satisfied for the universal
hypermultiplet). One can indeed easily check that the dualization in
Euclidean space changes the sign of the kinetic term. The conventions
are such that the matrices $M^{IJ}, \cG_{AB}$ and $A^I_A$ are the same
as in Minkowski space. Sign changes or factors of $\I$ upon Wick
rotation are never absorbed into these quantities, but written
explicitly.

In Lorentz signature, the complete supersymmetry transformation rules
were given in \cite{TV2}. For our purpose of calculating instanton
effects, it suffices to know the linearized supersymmetry transformation
rules. In Euclidean space, and for rigid supersymmetry, they can be
obtained by using \eqref{WickB} and \eqref{Wicksigma} in the Lorentzian
transformation rules:
 \begin{align}
  \de_\eps \phi^A & = \ga_{ia}^A\, \eps^i \la^a + \bar{\ga}^i{}^A_\ab\,
    \beps_i\, \bla^\ab\ \notag \\[2pt]
  \de_\eps B_{\mu\nu I} & = 2\I\, g_{Iia}\, \eps^i \si_{\mu\nu} \la^a
    - 2\I\, \bar{g}^i_{I\ab}\, \beps_i\, \bsi_{\mu\nu} \bla^\ab
    \notag \\[2pt]
  \de_\eps \la^a & = \big(\I \p_\mu \phi^A\, \W{A}{ai} + H_{\mu I}
    f^{Iai}\big)\, \si^\mu \beps_i + \dots \notag \\[2pt]
  \de_\eps \bla^\ab & = \big( \I \p_\mu \phi^A\, {\bar W}^\ab_{Ai} +
    H_{\mu I} \bar{f}^{I\ab}{}_i\big)\, \bsi^\mu \eps^i + \dots\ ,
    \label{susy-dtm}
 \end{align}
for some functions $\gamma^A_{ia}$, $g_{Iia}$ etc. The requirement of
closure of the supersymmetry algebra and invariance of the action
imposes constraints on and relations between the various quantities
appearing in the action and supersymmetry transformation rules. With
the parametrization given above, they are exactly the same as in
Lorentz signature \cite{TV2}, and we repeat them in appendix
\ref{sc-tens-rel}. The ellipsis stands for higher order terms in the
fermions that do not play a role at the linearized level. In comparison
with the Lorentzian case, the transformation rules for the bosonic
fields are the same, whereas for the fermions a factor of $\I$ appears
in the variation proportional to the tensor fields, as is consistent
with the rules of the Wick rotation.

\bigskip
\textbf{Local Supersymmetry} 
\medskip

We now couple the system to Euclidean $N=2$ supergravity. Pure $N=2$
supergravity in Euclidean space, without any matter couplings, was
constructed in \cite{TvN}. The scalar-tensor self-interactions in
Euclidean space were described above, and now we need its coupling to
the $N=2$ Poincar\'e supergravity multiplet, which contains a vielbein
$e_\mu{}^m$, the graviphoton $A_\mu$, and two gravitinos $\psi_\mu^i$,
$\bpsi_{\mu i}$, $i=1,2$. We follow the same Wick rotation rules as for
rigid supersymmetry, replace the metric by a Euclidean metric,
supplemented by
 \begin{equation}
  \psi_0^i \rightarrow \I \psi_4^i\ , \qquad \bpsi_{0i} \rightarrow
  \I \bpsi_{4i}\ ,\qquad A_0 \rightarrow \I A_4\ ,
 \end{equation}
for the gravitino and graviphoton. Notice again that in Euclidean
space $\psi^i_\mu$ and $\bpsi_{\mu i}$ are no longer related by
complex conjugation.

In Lorentz signature, the action and supersymmetry transformation rules
were given in \cite{TV2}, and after Wick rotation we get
 \begin{align} \label{EuclDTMaction}
  e^{-1} \cL & =  \frac{1}{2\kappa^2}\, R + \frac{1}{4}\, \cF^{\mu\nu}
    \cF_{\mu\nu} + \frac{1}{2}\, \cG_{AB}\, \hat{D}^\mu \phi^A\,
    \hat{D}_\mu \phi^B + \frac{1}{2}\, M^{IJ} \cH^\mu_I\,
    \cH_{\mu J} - \I A_A^I H^\mu_I\, \p_\mu \phi^A \notag \\
  & \tab + \I \ep^{\mu\nu\rho\si} (\cD_\mu \psi_\nu^i \si_\rho
    \bpsi_{\si i} + \psi_\si^i \si_\rho \cD_\mu \bpsi_{\nu i}) +
    \frac{\I}{2}\, h_{a\ab}\, (\la^a \si^\mu \cD_\mu \bla^\ab -
    \cD_\mu \la^a \si^\mu \bla^\ab) \notag \\
  & \tab - \kappa\, \cG_{AB} (\hat{D}_\mu \phi^A + \p_\mu \phi^A)\,
    (\ga^B_{ia}\, \la^a \si^{\mu\nu} \psi_\nu^i + \text{c.c.}) +
    \I \kappa M^{IJ} \cH^\mu_I\, (g_{Jia}\, \psi_\mu^i \la^a +
    \text{c.c.}) \notag \\[2pt]
  & \tab - \I \frac{\kappa}{2\2}\, (\tilde{\cF}^{\mu\nu} + \tilde{F}^{
    \mu\nu})\, (\psi_\mu^i \psi^{}_{\nu i} + \bpsi_\mu^i \bpsi^{}_{
    \nu i}) + \frac{\I\kappa}{2\2}\, \cF_{\mu\nu}\, (\cE_{ab}\,\la^a
    \si^{\mu\nu} \la^b - \text{c.c.}) \notag \\
  & \tab + \I M^{IJ} k_{Ja\ab}\, \la^a \si^\mu \bla^\ab \big[ \cH_{\mu
    I} + \I \kappa\, (g_{Iib}\, \psi_\mu^i \la^b + \text{c.c.})
    \big] \notag \\[4pt]
  & \tab + \kappa^2 M^{IJ} (g_{Iia}\, \psi_\mu^i \la^a + \text{c.c.})\,
    (g_{Jjb}\, \la^b \si^{\mu\nu} \psi_\nu^j + \text{c.c.})
    \notag \\
  & \tab + \frac{\kappa^2}{8}\, (\cE_{ac} \cE_{bd}\, \la^a \la^b\ \la^c
    \la^d + \text{c.c.}) - \frac{1}{4}\, V_{ab\,\ab\bb}\, \la^a
    \la^b\, \bla^\ab \bla^\bb\ .
 \end{align}
By ``c.c.'' above, and further on below, we mean the analytic
continuation of the complex conjugated expressions in Lorentz
signature. The covariant derivatives $\cD_\mu$ of the fermions contain
connections $\G{A}{a}{b}$ and $\G{A}{i}{j}$, just in the same way as in
Minkowski space. We refer to \cite{TV2} for more details about the
connections, supercovariant derivatives and field strengths. Notice also
the appearance of a new four-fermi term proportional to the tensor
$\cE_{ab}$ as defined in \eqref{def-E-tensor}.

It is important to mention here that we are working in a 1.5 order
formalism, in which the spin-connection is determined by its own field
equation. Hence, it contains gravitinos as well as matter fermion
bilinears. This formalism simplifies checking supersymmetry of the
action, but hides certain quartic fermion terms, both in the gravitino
and in the matter sector. One would therefore have to be careful in
interpreting the calculation of instanton corrections to four-fermi
correlators, which would be nonzero as a result of the zero mode
counting. Fortunately, as we will see later on, there is a more
convenient way that avoids this source of confusion.

The Euclidean linearized supersymmetry transformation rules are again
given by \eqref{susy-dtm}, where all derivatives and field strengths are
replaced by supercovariant ones, and \cite{TV2}
\begin{equation} \label{de_B_loc}
 \de_\eps B_{\mu\nu I} = 2\I\, g_{Iia}\, \eps^i \si_{\mu\nu} \la^a
 - 4 \kappa^{-1} {\Omega_I}^i{}_j\, \eps^j \si_{[\mu} \bpsi_{\nu]i}
 + \text{c.c.}\ .
\end{equation}
The transformations of the supergravity multiplet are given by
 \begin{align}
  \de_\eps e_\mu{}^m & = \I \kappa\, (\eps^i \si^m \bpsi^{}_{\mu i} -
    \psi_\mu^i \si^m \beps_i) \notag \\[2pt]
  \de_\eps A_\mu & = \I\2\, (\eps_i \psi_\mu^i + \beps^i \bpsi_{\mu
    i}) \notag \\[2pt]
  \de_\eps \psi_\mu^i & = \kappa^{-1}\, \cD_\mu \eps^i + \frac{1}{\2}\,
    \ep^{ij} F^+_{\mu\nu}\, \si^\nu \beps_j - \I \kappa^{-1}
    H_{\mu I} \Gamma^{Ii}{}_j\, \eps^j + \dots \notag \\[2pt]
 \de_\eps \bpsi_{\mu i} & = \kappa^{-1}\, \cD_\mu \beps_i + \frac{1}
    {\2}\, \ep_{ij} F^-_{\mu\nu}\, \bsi^\nu \eps^j + \I \kappa^{-1}
    H_{\mu I} \Gamma^{Ij}{}_i\, \beps_j + \dots\ ,
 \end{align}
where in the last two lines we have denoted the (anti-) selfdual
graviphoton field strengths by $F_{\mu\nu}^{\pm}=\half(F_{\mu\nu}\pm
\tilde{F}_{\mu\nu})$, and we have dropped fermion bilinears. Just as for
the rigid case, they can easily be reinserted.

The coefficient functions $g_{Iia}$, $\W{A}{ai}$, etc.\ appearing in the
above equations are related to hypermultiplet quantities in the same way
as in the rigid case. Hence, they satisfy the same relations
\eqref{rel1}--\eqref{EWidentity}. Moreover, we have the relation
 \begin{equation} \label{Gamma-Omega}
  \Gamma^{Ii}{}_j = M^{IJ} {\Omega_J}^i{}_j\ ,
 \end{equation}
between the coefficients which appear in the supersymmetry
transformations of the gra\-vi\-tinos and tensors, respectively.

\section{The supersymmetric double-tensor multiplet} \label{sect_DTM}

The purpose of the previous section was twofold: to give the generic
form of the low-energy effective action, and to discuss the
supersymmetry rules that will be needed to study the BPS properties of
the fivebrane instantons that appear in the (classical) double-tensor
multiplet action \eqref{DTM-action}. This multiplet and its action are
a specific example of the general case given in \eqref{EuclDTMaction},
for which\footnote{From now on we set again $\kappa^{-2}=2$ and rescale
the supersymmetry parameters $\eps^i$, $\beps_i$ by a factor of $\2$.
\label{ftnote3}}
 \begin{equation} \label{DTM_metrics}
  \cG_{AB} = \lp 1 & 0 \\[2pt] 0 & \e^{-\phi} \rp\ ,\qquad M^{IJ} =
  \e^{\phi} \lp 1 & -\chi \\[2pt] -\chi & \e^\phi + \chi^2 \rp\ ,\qquad
  A_A^I = 0\ .
 \end{equation}
Strictly speaking, instead of $A^I_A=0$ we could allow for a
nonvanishing connection with trivial field strength $F_{AB}{}^I=2
\p^{}_{[A}A_{B]}^I=0$. Such connections are pure gauge and lead to
total derivatives in the action. In perturbation theory, they can be
dropped, but nonperturbatively they can be nonvanishing and lead to
imaginary theta-angle-like terms. We have discussed and included such
terms seperately in \eqref{theta-angles} and \eqref{theta-angles2}, so
it suffices to set $A_A^I=0$. The complete double-tensor multiplet,
including all other coefficient functions that determine the action
and supersymmetry transformations, was written down in \cite{TV1}, and
is summarized in appendix \ref{coeff_DTM}.

Using these results, we can now give the linearized (in the fermions)
Euclidean supersymmetry transformations of the fermions. They can be
written as
 \begin{alignat}{2}
  \de_\eps \la^a & = \I\2\, E_\mu^{ai}\, \si^\mu \beps_i\ , &\qquad
    \de_\eps \bpsi_{\mu i} & = 2 \bar{D}_{\mu\,i}{}^j\, \beps_j +
    \ep_{ij} F^-_{\mu\nu}\, \bsi^\nu \eps^j \notag \\[2pt]
  \de_\eps \bla^\ab & = \I\2\, \bar{E}_{\mu\,i}^\ab\, \bsi^\mu \eps^i\
    , &\qquad \de_\eps \psi_\mu^i & = 2 {D_\mu}^i{}_j\, \eps^j +
    \ep^{ij} F^+_{\mu\nu}\, \si^\nu \beps_j\ , \label{susyMN}
 \end{alignat}
where we have introduced
 \begin{alignat}{2} \label{defMN}
  E_\mu^{ai} & = \p_\mu \phi^A W_{\!A}^{ai} - \I H_{\mu I} f^{Iai}\ ,
    &\qquad \bar{D}_{\mu\,i}{}^j & = \de^j_i \nabla_{\!\mu} - \p_\mu
    \phi^A \Gamma_{\!A}{}^j{}_i + \I H_{\mu I} \Gamma^{Ij}{}_i
    \notag \\[2pt]
  \bar{E}_{\mu\,i}^\ab & = \p_\mu \phi^A \bar{W}_{\!Ai}^\ab - \I
    H_{\mu I} \bar{f}^{I\ab}{}_i\ , &\qquad {D_\mu}^i{}_j & =
    \de^i_j \nabla_{\!\mu} + \p_\mu \phi^A \Gamma_{\!A}{}^i{}_j -
    \I H_{\mu I} \Gamma^{Ii}{}_j\ ,
 \end{alignat}
with $\nabla_{\!\mu}$ the Lorentz-covariant derivative. We will find
useful the observation that $\bar{E}_\mu$ and $D_\mu$ are related to
their counterparts $E_\mu$ and $\bar{D}_\mu$ according
to\footnote{Note that in the second identity the covariant derivatives
in $D_\mu$ and $\bar{D}_\mu$ are in the same representation of Spin(4),
whereas in \eqref{susyMN} they are not.}
 \begin{equation} \label{MNrel}
  \bar{E}_{\mu\,j}^\ab = - h^{\ab a} \cE_{ab}\, E_\mu^{bl}\, \ep_{lj}\
  ,\qquad {D_\mu}^i{}_j = \ep^{ik} \bar{D}_{\mu\,k}{}^l\, \ep_{lj}\ .
 \end{equation}
The first identity is due to the relation \eqref{EWidentity}, while
the second is a consequence of $\SU(2)$-covariant constancy of
$\ep_{ij}$.

More explicitly, we have, at the linearized level,
 \begin{align}
  \de_\eps \lp \la^1 \\[2pt] \la^2 \rp & = \I \lp \e^{-\phi/2} \p_\mu
    \chi - \e^{\phi/2} \hat{H}_{\mu1} & \p_\nu \phi + \e^\phi
    H_{\nu2} \\[2pt] - \p_\mu \phi + \e^\phi H_{\mu2} & \e^{-\phi/2}
    \p_\nu \chi + \e^{\phi/2} \hat{H}_{\nu1} \rp \lp \si^\mu
    \beps_1 \\[2pt] \si^\nu \beps_2 \rp \notag \\[4pt]
  \de_\eps \lp \bla^1 \\[2pt] \bla^2 \rp & = \I \lp \e^{-\phi/2} \p_\mu
    \chi + \e^{\phi/2} \hat{H}_{\mu1} & \p_\nu \phi - \e^\phi
    H_{\nu2} \\[2pt] - \p_\mu \phi - \e^\phi H_{\mu2} & \e^{-\phi/2}
    \p_\nu \chi - \e^{\phi/2} \hat{H}_{\nu1} \rp \lp \bsi^\mu \eps^1
    \\[2pt] \bsi^\nu \eps^2 \rp \label{ferm-transf}
 \end{align}
for the matter fermions, and
 \begin{align}
  \de_\eps \lp \psi_\mu^1 \\[2pt] \psi_\mu^2 \rp & = \lp 2 \nabla_{\!
    \mu} + \half \e^\phi H_{\mu2} & - \e^{-\phi/2} \p_\mu \chi +
    \e^{\phi/2} \hat{H}_{\mu1} \\[2pt] \e^{-\phi/2} \p_\mu \chi +
    \e^{\phi/2} \hat{H}_{\mu1} & 2 \nabla_{\!\mu} - \half \e^\phi
    H_{\mu2} \rp \lp \eps^1 \\[2pt] \eps^2 \rp + \dots \notag
    \\[4pt]
  \de_\eps \lp \bpsi_{\mu1} \\[2pt] \bpsi_{\mu2} \rp & = \lp 2
    \nabla_{\!\mu} - \half \e^\phi H_{\mu2} & - \e^{-\phi/2} \p_\mu
    \chi - \e^{\phi/2} \hat{H}_{\mu1} \\[2pt] \e^{-\phi/2} \p_\mu
    \chi - \e^{\phi/2} \hat{H}_{\mu1} & 2 \nabla_{\!\mu} + \half
    \e^\phi H_{\mu2} \rp \lp \beps_1 \\[2pt] \beps_2 \rp + \dots
    \label{gravitino-trans}
 \end{align}
for the gravitinos, where we have omitted the graviphoton terms.

We end this section by giving the fermionic equations of motion, at the
linearized level. For the hyperinos we find
 \begin{align}
  & \I \si^\mu \cD_\mu \bla^\ab + H_{\mu I}\, \bar{\Gamma}^{I\ab}
    {}_{\bb}\, \si^\mu \bla^\bb + \frac{\I}{2}\, h^{\ab a}
    \cE_{ab}\, F^+_{\mu\nu}\, \si^{\mu\nu} \lambda^b = - \frac{1}
    {\2}\, \si^\nu \bar{E}^{\ab}_{\mu\,i} \bsi^\mu \psi_\nu^i
    \label{hyperino-EOM} \\[2pt]
  & \I \bsi^\mu \cD_\mu \la^a + H_{\mu I}\, \Gamma^{Ia}{}_{b}\, \bsi^\mu
    \la^b - \frac{\I}{2}\, h^{\ab a} \bar{\cE}_{\ab\bb}\, F^-_{\mu
    \nu}\, \bsi^{\mu\nu} \bla^\bb = -\frac{1}{\2}\, \bsi^\nu
    {E}^{ai}_{\mu} \si^\mu \bpsi_{\nu\,i}\ .
 \end{align}
What makes these different from the usual Dirac-like equation is the
presence of the mixing term with the (anti-) selfdual graviphoton field
strenth and the inhomogeneous gravitino term originating from its
coupling to the rigid supersymmetry current of the double-tensor
multiplet. This will become important in the discussion of the fermionic
zero modes.

The gravitino field equations read
 \begin{align}
  & \I \ep^{\mu\nu\rho\si} \si_\rho \cD_\si \bpsi_{\nu\,i} - \ep^{\mu
    \nu\rho\si} H_{\si I} \Gamma^{Ij}{}_i\, \si_\rho \bpsi_{\nu j}
    - \I F^{\mu\nu-} \psi_{\nu\,i} = \frac{1}{2\2}\, h_{a\ab}
    \bar{E}^{\ab}_{\nu\,i}\, \si^\nu \bsi^\mu \lambda^a
    \label{gravitino-EOM} \\[2pt]
  & \I \ep^{\mu\nu\rho\si} \bsi_\rho \cD_\si \psi_\nu^i + \ep^{\mu\nu
    \rho\si} H_{\si I} \Gamma^{Ii}{}_j\, \bsi_\rho \psi_\nu^j + \I
    F^{\mu\nu+} \bpsi_\nu^i = - \frac{1}{2\2}\, h_{a\ab} E^{ai}_{
    \nu}\, \bsi^\nu \si^\mu \bla^\ab\ . \label{gravitino-EOM2}
 \end{align}
Notice that one can combine the first two terms on the left-hand side
into the operator ${D_\mu}^i{}_j$, as defined in \eqref{defMN}.

The fermionic field equations will be important for finding the
fermionic zero modes. We return to this in section \ref{sect_FZM}.

\section{Unbroken supersymmetries}
\label{sect_unbroken}

We now determine the supersymmetries left unbroken by our instanton
solutions. These are generated by Killing spinors $\eps^i$, $\beps_i$
which give vanishing supersymmetry transformations \eqref{susyMN} of the
fermions in the bosonic instanton background. We shall concentrate on
the parameters $\beps_i$ in the following, the $\eps^i$ can be obtained
from these as explained below.

We begin by writing the $\la^a$ transformations as
 \begin{align}
  \de_\eps \la^1 & = - 2\I\, \si^\mu (\nabla_{\!\mu} - \half \p_\mu
    \phi - \quart \e^\phi H_{\mu2}) \beps_2 + \I \si^\mu \de_\eps
    \bpsi_{\mu2} \notag \\[2pt]
  \de_\eps \la^2 & = 2\I\, \si^\mu (\nabla_{\!\mu} - \half \p_\mu \phi
    + \quart \e^\phi H_{\mu2}) \beps_1 - \I \si^\mu \de_\eps
    \bpsi_{\mu1}\ ,
 \end{align}
which holds identically for any purely bosonic field configuration with
$F_{\mu\nu}^-=0$. Requiring the transformations of the fermions to
vanish imposes the necessary conditions
 \begin{align}
  \si^\mu (\nabla_{\!\mu} - \half \p_\mu \phi + \quart \e^\phi
    H_{\mu2})\, \beps_1 & = 0 \notag \\[2pt]
  \si^\mu (\nabla_{\!\mu} - \half \p_\mu \phi - \quart \e^\phi
    H_{\mu2})\, \beps_2 & = 0\ . \label{Killing}
 \end{align}
They simplify upon using the Bogomol'nyi condition. Eq.~\eqref{O2BPS}
implies that $\de_\eps\la^1=\de_\eps\bla^2=0$ identically for instantons
($\de_\eps\la^2=\de_\eps\bla^1=0$ for anti-instantons). Accordingly, the
solutions $\beps_i$ to \eqref{Killing} will also solve $\si^\mu\de_\eps
\bpsi_{\mu2}=0$. Using the Bogomol'nyi condition for $H_{\mu2}$ in
\eqref{Killing}, we can write the $\beps_i$ as
 \begin{equation}
  \beps_1 = \e^{(2\pm1)\phi/4}\, \bar{\eta}_1\ ,\qquad \beps_2 =
  \e^{(2\mp1)\phi/4}\, \bar{\eta}_2\ ,
 \end{equation}
where the spinors $\bar\eta_i$ have to satisfy $\slsh{\nabla}\bar\eta_i
=0$. We consider only flat space as a background in this paper, and zero
modes of the operator $\p\!\!\!/$ will be discussed in detail in the
next section. We shall find there that they are of the form $\bar{\eta}
=\mathrm{const}+\p_\mu h\,\bsi^\mu\xi$, where $h(x)$ is a harmonic
function and $\xi$ a constant spinor. For instantons the condition
 \begin{equation}
  \half \de_\eps \bpsi_{\mu2} = (\p_\mu - \quart \p_\mu \phi)\, \beps_2
  = \e^{\phi/4}\, \p_\mu \bar{\eta}_2 = 0
 \end{equation}
leaves only $\bar\eta_2=\mathrm{const}$. It remains to consider the
transformation of $\bpsi_{\mu1}$; using the Bogomol'nyi condition for
both tensors and the above expressions for $\beps_i$ yields
 \begin{equation}
  \half \de_\eps \bpsi_{\mu1} = (\p_\mu + \quart \p_\mu \phi)\, \beps_1
  - \e^{-\phi/2} \p_\mu \chi\, \beps_2 = \e^{-\phi/4}\, \p_\mu (\e^\phi
  \bar{\eta}_1 - \chi \bar{\eta}_2) = 0\ .
 \end{equation}
Using the fact that $\bar\eta_2=\mathrm{const}$ and $\bar\eta_1=
\mathrm{const}+\p_\mu h\,\bsi^\mu\xi$ leads to two possibilities. For
$\bar\eta_1=\mathrm{const}$ it follows that $\bar\eta_1=\rho\,\bar
\eta_2$ and
 \begin{equation} \label{chi_constr}
  \chi = \rho\, \e^\phi + \chi_0\ ,
 \end{equation}
where $\chi_0$, $\rho$ are constants. Using the notation introduced in
section \ref{Fivebrane-inst}, we can rewrite $\rho=g_s^2\Delta\chi$.
\eqref{chi_constr} constrains the instanton configuration beyond the
restrictions imposed by the Bogomol'nyi condition. As explained in
section \ref{Fivebrane-inst}, it requires that all $\chi_i$ are equal.
Note that this relation between $\chi$ and $\phi$ follows automatically
if one imposes spherical symmetry, since then the harmonic function
$\e^{-\phi} \chi$ must depend linearly on the harmonic function
$\e^{-\phi}$ \cite{TV1}.

For $\bar\eta_1=\p_\mu h\,\bsi^\mu\xi$, we did not find any non-trivial
solutions.

The same analysis can be done for anti-instantons. In either case $\bar
\eta_1$ turns out to be proportional to $\bar\eta_2$, so we conclude
that the unbroken supersymmetries $\beps_i$ are given in terms of one
constant spinor $\bar\eta$ as $\beps_i(x)=u_i(x)\,\bar\eta$, with
functions
 \begin{equation}
  \lp u_1 \\[2pt] u_2 \rp = \e^{\phi/2} \lp \rho^{(1\pm1)/2}\, \e^{\pm
  \phi/4} \\[2pt] \pm \rho^{(1\mp1)/2}\, \e^{\mp\phi/4} \rp\ .
 \end{equation}
The Killing spinors of opposite chirality are then given by $\eps^i=
\ep^{ij}u_j\,\eta$, with $\eta$ another arbitrary constant spinor. This
immediately follows from the relations \eqref{MNrel}: if $u_i$ are
(spinless) zero modes of $E_\mu$ and $\bar{D}_\mu$, then $\ep^{ij}u_j$
are zero modes of $\bar{E}_\mu$ and $D_\mu$. We conclude that the
fivebrane instanton in flat space, subject to the additional constraint
\eqref{chi_constr}, preserves one half of the
supersymmetries\footnote{The trivial solution with both $\e^{-\phi}$
and $\chi$ constants of course preserves all the supersymmetries. Such
solutions are not instantons but rather, they parametrize the vacua
which the instantons interpolate between.}.

\section{Fermionic zero modes and broken supersymmetries}
\label{sect_FZM}

In this section we determine the fermionic zero modes and show that all
of them can be obtained by acting with the broken supersymmetries on the
purely bosonic solution. We first concentrate on instantons; the case of
anti-instantons is similar and will be summarized at the end of this
section.

\bigskip
\textbf{Spin 1/2 zero modes} 
\medskip

The zero modes for the matter fermions are given by the solutions to the
linearized hyperino equations of motion \eqref{hyperino-EOM}, which are
coupled to the gravitinos. For the double-tensor multiplet we have
$\Gamma_{\!A}{}^a{}_b=\Gamma^{1\,a}{}_b=0$; plugging in the only
nonvanishing coefficient $\Gamma^{2\,a}{}_b$ given in \eqref{Gamma2ab}
and using the Bogomol'nyi condition \eqref{O2BPS} (with the upper sign)
for $H_{\mu2}$, the two hyperino equations become
 \begin{equation} \label{5br-zeromode-eqn1}
  \I \si^\mu \lp \p_\mu \bla^1 - \tfrac{3}{4} \p_\mu \phi\, \bla^1
  \\[3pt] \p_\mu \bla^2 + \tfrac{3}{4} \p_\mu \phi\, \bla^2 \rp = -
  \si^\nu \bsi^\mu \lp \e^{-\phi/2} \p_\mu \chi\, \psi_\nu^1 + \p_\mu
  \phi\, \psi_\nu^2 \\[3pt] 0 \rp\ .
 \end{equation}
Notice that the $\bla^2$ equation has no coupling to the gravitinos.
Similarly for the unbarred hyperinos,
 \begin{equation} \label{5br-zeromode-eqn2}
  \I \bsi^\mu \lp \p_\mu \la^1 + \tfrac{3}{4} \p_\mu \phi\, \la^1
  \\[3pt] \p_\mu \la^2 - \tfrac{3}{4} \p_\mu \phi\, \la^2 \rp = \bsi^\nu
  \si^\mu \lp 0 \\[3pt] \p_\mu \phi\, \bpsi_{\nu1} -\e^{-\phi/2} \p_\mu
  \chi\, \bpsi_{\nu2} \rp\ ,
 \end{equation}
but now $\la^1$ decouples from the gravitinos.

It is useful to first study zero modes of the operator
 \begin{equation}\label{Dk}
  \slsh{D}_k \equiv \sigma^\mu (\p_\mu - k \p_\mu \phi) = \e^{k\phi}
  \p\!\!\!/\, \e^{-k\phi}\ ,\qquad k \in \fieldR\ .
 \end{equation}
In the absence of gravitinos, this is precisely the relevant zero mode
operator. Clearly, the zero modes of $\slsh{D}_k$ are in one-to-one
correspondence with those of $\p\!\!\!/$. But whereas solutions $\bar
\zeta$ to $\p\!\!\!/\,\bar\zeta=0$ are not normalizable, the
corresponding modes $\bla=\e^{k\phi}\bar\zeta$ of $\slsh{D}_k$ can be
normalizable for appropriate values of $k$. In flat Euclidean space,
with the origin $x_0$ not excised, the only solution for $\bar\zeta$ is
a constant spinor. However, when the origin is cut out, there is another
non-trivial solution:
 \begin{equation}
  \bar{\zeta}(x) =  2\I\, \p_\mu h(x)\, \bsi^\mu \xi\ ,
 \end{equation}
where $\xi$ is a constant spinor, $h$ is a harmonic function, and the
factor $2\I$ is a choice of normalization. This is the only solution, as
one can show by rewriting the two-component spinor as $\bar{\zeta}(x)=
2\I\,f_\mu(x)\,\bsi^\mu\xi$ for arbitrary real functions $f_\mu$; the
equation $\p\!\!\!/\,\bar\zeta=0$ then imposes the conditions $\p_{[\mu}
f_{\nu]}=\p_\mu f^\mu=0$. The constant solution cannot lead to a
normalizable zero mode\footnote{Normalizable zero modes $Z$ must
satisfy $\int_0^\infty\mathrm{d}r\,r^3 |Z|^2<\infty$. This implies that
the asymptotic behavior of $Z$ at infinity must go to zero like
$r^{-5/2}$ or faster, and at the origin $Z$ may not diverge faster than
$r^{-3/2}$.} and must therefore be discarded. The possibly normalizable
solutions to $\slsh{D}_k\bla=0$ are then given by
 \begin{equation}
  \bla = 2\I\, \e^{k\phi}\, \p_\mu h\, \bsi^\mu \xi\ ,
 \end{equation}
with $\xi$ a constant spinor. For spherical symmetry, the only harmonic
function available is $\e^{-\phi}$. Looking at the asymptotic behavior,
one finds that the only normalizable zero modes are
 \begin{equation} \label{zm-example}
  \bla = 2\I\, \e^{k\phi}\, \p_\mu \e^{-\phi}\, \bsi^\mu \xi\ ,\qquad k
  \geq \tfrac{3}{4}\ .
 \end{equation}
Comparing with the hyperino zero mode equations
\eqref{5br-zeromode-eqn1} and \eqref{5br-zeromode-eqn2}, we conclude
that
 \begin{equation} \label{nozeromodes}
  \bla^2 = 0\ ,\qquad \la^1 = 0\ ,
 \end{equation}
since $k=-3/4$. Zero modes for $\bla^1$ and $\la^2$ would look like
\eqref{zm-example} with the smallest possible value for $k$ if no
gravitinos were present. Before proceeding, we will determine the
gravitinos from their own field equations. The solution we then plug
into \eqref{5br-zeromode-eqn1}, \eqref{5br-zeromode-eqn2} and solve for
$\bla^1$ and $\la^2$.

\bigskip
\textbf{No spin 3/2 zero modes} 
\medskip

In the instanton background, \eqref{gravitino-EOM} becomes
 \begin{equation}
  \I \ep^{\mu\nu\rho\si} \si_\rho \lp \p_\si \bpsi_{\nu1} + \tfrac{1}{4}
  \p_\si \phi\, \bpsi_{\nu1} - \e^{-\phi/2} \p_\si \chi\, \bpsi_{\nu2}
  \\[3pt] \p_\si \bpsi_{\nu2} - \tfrac{1}{4} \p_\si \phi\, \bpsi_{\nu2}
  \rp = \frac{1}{2} \lp \e^{-\phi/2} \p_\nu \chi \\[3pt] \p_\nu \phi \rp
  \si^\nu \bsi^\mu \la^1\ .
 \end{equation}
Notice that only $\la^1$ couples to the gravitino; this is consistent
with the hyperino equations of motion as one can easily check. But we
have just concluded in \eqref{nozeromodes} that $\la^1$ vanishes in the
instanton background, i.e.\ there is no zero mode in $\la^1$. This means
that the inhomogeneous term on the right-hand side of the gravitino
equations vanishes, and that the second gravitino decouples from the
first. Writing
 \begin{equation}
  \bpsi_{\mu2} = \e^{\phi/4} \bar{\zeta}_{\mu2}\ ,
 \end{equation}
we can, using some sigma matrix identities such as \eqref{eps-sigma},
rewrite the second gravitino equation of motion as
 \begin{equation} \label{zeta-mode}
  \si^\mu (\p_\mu\, \bar{\zeta}_{\nu2}- \p_\nu\, \bar{\zeta}_{\mu2})
  = 0\ .
 \end{equation}
Of course, there are an infinite number of solutions to this equation,
for example $\bar{\zeta}_\mu=\p_\mu\bar{\zeta}$, but we still have to
subject them to the supersymmetry gauge-fixing procedure. We adopt the
common gauge
 \begin{equation} \label{gaugefix}
  \bsi^\mu \psi_\mu^i = \si^\mu \bpsi_{\mu i} = 0\ ,
 \end{equation}
in particular $\si^\mu\bar{\zeta}_{\mu2}=0$. Then, \eqref{zeta-mode}
further simplifies to $\p\!\!\!/\,\bar{\zeta}_{\mu2}=0$, still subject
to \eqref{gaugefix}. The solution is given by
 \begin{equation} \label{zeta-mu-mode}
  \bar{\zeta}_{\mu2}= \p_\mu \p_\nu h \,{\bar \sigma}^\nu \xi\ ,
 \end{equation}
with $\xi$ a constant spinor and $h$ a harmonic function. For a given
solution, one can still act with residual supersymmetry transformations
that preserve the gauge \eqref{gaugefix}. The gravitino supersymmetry
transformations in the NS5-brane instanton (upper sign in \eqref{O2BPS})
background read
 \begin{align}
  \de_\eps \bpsi_{\mu 1} & = 2 (\p_\mu + \quart \p_\mu \phi) \beps_1 -
    2 \e^{-\phi/2}\, \p_\mu \chi\, \beps_2 \notag \\[2pt]
  \de_\eps \bpsi_{\mu 2} & = 2 (\p_\mu - \quart \p_\mu \phi) \beps_2\ .
 \end{align}
Such a supersymmetry transformation has the effect of shifting
$\bar{\zeta}_{\mu2}$ with a total derivative, so it obviously still
satisfies \eqref{zeta-mode}. Imposing furthermore the gauge-fixing
condition implies that the residual supersymmetry transformations must
satisfy
 \begin{equation} \label{res-susy}
  \p\!\!\!/\, (\e^{-\phi/4} {\bar \eps_2})=0\ .
 \end{equation}
The solutions are again given by a constant or by the derivative of a
harmonic function and acting with such a residual supersymmetry
transformation remains inside the class given by \eqref{zeta-mu-mode}.
Therefore the only candidate gravitino zero mode for $\bpsi_{\mu2}$ is
given by
 \begin{equation}
  \bpsi_{\mu2} = \e^{\phi/4} \p_\mu \p_\nu h\, \bsi^\nu \xi\ ,
 \end{equation}
where $h$ can again be chosen to be $h=\e^{-\phi}$ for spherically
symmetric harmonic functions. However, this zero mode is not
normalizable, since at the origin it diverges too fast, as one can
explicitly check. As a consequence, there is no normalizable gravitino
zero mode for $\bpsi_{\mu2}$, and we can set it to zero in the equation
of motion for the first gravitino. A similar analysis now shows that
there is no normalizable zero mode for $\bpsi_{\mu1}$ either, and so we
can set both gravitinos to zero in the hyperino equations for $\la^2$
and $\bla^1$. The hyperino zero modes then follow from the discussion
above.

\bigskip
\textbf{Index theorems?} 
\medskip

We conclude that the solutions of the linearized fermion equations of
motion in the presence of the fivebrane instanton are given by
 \begin{equation}
  \bpsi_{\mu i} = \psi_\mu^i = \la^1 = \bla^2 = 0\ ,
  \end{equation}
and the four zero modes lie in
 \begin{equation} \label{lambda_zm}
  \bla^1 = 2\I\,\e^{3\phi/4}\, \p_\mu \e^{-\phi}\, \bsi^\mu \xi\ ,\qquad
  \la^2= 2\I\,\e^{3\phi/4}\, \p_\mu \e^{-\phi}\, \si^\mu \bxi\ ,
 \end{equation}
where the four fermionic collective coordinates are denoted by the two
unrelated spinors $\xi$ and $\bxi$. Notice the difference with
Yang-Mills theories, where fermionic zero modes only appear in one
chiral sector.

These four fermionic zero modes should be counted by the index of the
operators
 \begin{equation}
  \slsh{D}_k \equiv \e^{k\phi} \p\!\!\!/\, \e^{-k\phi}\ ,\qquad
  \slsh{\bar D}_k \equiv \e^{k\phi} \bar{\p}\!\!\!/\, \e^{-k\phi}\
  ,\qquad k \in \fieldR\ ,
 \end{equation}
where $\p\!\!\!/=\si^\mu\p_\mu$ and $\bar{\p}\!\!\!/=\bsi^\mu\p_\mu=
\p\!\!\!/^\dagger$, see \eqref{Dk}. Clearly, these operators are not
hermitean, but satisfy $(\slsh{D}_k)^\dagger=\slsh{\bar D}_{\!-k}$.
Due to this property, they are different from the usual Dirac operators
with anti-hermitean connection. The case we are interested in
corresponds to $k=3/4$. The relevant indices are defined as
 \begin{equation}
  \Ind \slsh{D}_k \equiv \dim \ker \slsh{D}_k - \dim \ker \slsh{\bar
  D}_{\!-k}\ ,\qquad \Ind \slsh{\bar D}_k \equiv \dim \ker \slsh{\bar
  D}_k - \dim \ker \slsh{D}_{\!-k}\ .
 \end{equation}
In the beginning of this section we have shown explicitly that both
$\slsh{D}_{\!-3/4}$ and $\slsh{\bar D}_{\!-3/4}$ have no normalizable
zero modes, so the index for $k=3/4$ indeed counts the number of zero
modes. Index theorems for our kind of operators and spacetime topology
should therefore reproduce that $\Ind\slsh{\bar D}_{3/4}=\Ind
\slsh{D}_{3/4}=2$. It would be interesting to find and apply such index
theorems to instanton solutions in scalar-tensor theories in general.

\bigskip
\textbf{Broken supersymmetries} 
\medskip

We now show that these zero modes are precisely generated by acting with
the residual supersymmetries \eqref{res-susy} on the hyperinos. As
already said, the solutions for $\e^{-\phi/4}\beps_2$ are either
constants or given in terms of a harmonic function. The latter lead to
non-normalizable gravitino zero modes which we have discarded, but the
constant spinor solution leaves the second gravitino invariant,
 \begin{equation} \label{broken-susy2}
  \beps_2 = \e^{\phi/4}\, \bar{\eta}\ ,
 \end{equation}
with $\p_\mu\bar{\eta}=0$. The broken supersymmetries for $\beps_1$ are
then determined by those transformations that leave $\bpsi_{\mu1}$
invariant and preserve the gauge condition \eqref{gaugefix}, but which
act nontrivially on the spin 1/2 fermions. Using \eqref{broken-susy2}
we find
 \begin{equation} \label{broken-susy1}
  \beps_1 = \e^{-\phi/4} \big( \chi \bar{\eta} + \bxi' \big)\ ,
 \end{equation}
where $\bxi'$ is another constant spinor. With the $\beps_i$ inserted
into the $\la^a$ transformations, we find that, after a redefinition
$\bxi=\bxi'+\chi_0\bar\eta$, the Killing spinor proportional to
$\bar\eta$ leaves the fields invariant and thus generates unbroken
supersymmetries, as observed in section \ref{sect_unbroken}. The
Killing spinor proportional to $\bxi$ on the other hand yields the
$\la$-zero mode\footnote{The left superscript counts the Grassmann
collective coordinates (GCC) in the fields (excluding the supersymmetry
parameters). We omit the superscript ${}^{(0)}$.}
 \begin{equation} \label{O2ZM_xi}
  {}^{(1)}\!\la^1 = 0\ ,\qquad {}^{(1)}\!\la^2 = -2\I\, \e^{-\phi/4}\,
  \p_\mu \phi\, \si^\mu \bxi\ .
 \end{equation}
Using the relation between Killing spinors of opposite chirality
derived in the previous section, we immediately obtain the unbarred
broken supersymmetries:
 \begin{equation}
  \eps^1 = 0\ ,\qquad \eps^2 = -\e^{-\phi/4} \xi\ .
 \end{equation}
Here $\xi$ is another constant spinor. The zero mode generated by
$\eps^2$ is found to be
 \begin{equation}
  {}^{(1)}\!\bla^1 = -2\I\, \e^{-\phi/4}\, \p_\mu \phi\, \bsi^\mu \xi\
  ,\qquad  {}^{(1)}\!\bla^2 = 0\ .
 \end{equation}
Notice that this is a function of space $x$ and of the position $x_0$
of the instanton. One can diagrammatically represent this fermionic zero
mode by a line connecting the fermion at $x$ to the instanton at the
point $x_0$, cf.\ \eqref{ferm_prop}.

Similar results can be obtained for the anti-instanton, the zero modes
will now be in $\la^1$ and $\bla^2$. For further convenience below, we
introduce a handy notation that connects the hyperino-labels ``1'' and
``2'' with the (anti-) instanton-labels ``$+$'' and ``$-$''. This is
done in such a way that the hyperino labels 1 and 2 are denoted by upper
and lower indices respectively. These indices can then be further
specified by indicating the background, instanton or anti-instanton. In
this notation, the absence of fermionic zero modes is expressed by the
equations ${}^{(1)}\!\la^\pm=0$, where the upper index is associated to
the first hyperino in the instanton ($+$) background, and the lower
label is associated to the second hyperino in the anti-instanton ($-$)
background. Similarly we have that ${}^{(1)}\!\bla^\mp=0$. For the
broken supersymmetries we have $\eps^\pm=\beps_\mp=0$. In the zero mode
sector we can write
 \begin{equation} \label{summ-ferm-zeromodes}
  {}^{(1)}\la^\mp = -2\I\, \e^{-\phi/4}\, \p_\mu \phi\, \si^\mu
  \bxi^\mp\ ,\qquad {}^{(1)}\bla^\pm = \mp 2\I\, \e^{-\phi/4}\, \p_\mu
  \phi\, \bsi^\mu \xi^\pm\ .
 \end{equation}
The fermionic collective coordinates $\bxi^\mp$ are two independent
two-component spinors which distinguish between instantons and
anti-instantons. Similarly for $\xi^\pm$, and we have have put in an
additional sign for the anti-instanton fermionic zero mode in
$\bla^2$ for later convenience. For the broken supersymmetries, we have
 \begin{equation}
  \eps^\mp = - \e^{-\phi/4}\, \xi^\mp\ ,\qquad {\bar\eps}_\pm = \pm
  \e^{-\phi/4}\, \bxi^\pm\ .
 \end{equation}
In further sections, we sometimes drop these $\pm$ indices on $\xi$ and
$\bxi$, to avoid heavy notation and because it is always clear from the
context what is meant.

\section{Measure on the instanton moduli space} \label{sect_measure}

{}From field theory instanton calculations, we learn that the path
integral measure reduces to a finite dimensional integral over the
collective coordinates (the instanton moduli space), together with a
path integral over the quantum fluctuations around the instanton. The
measure on the instanton moduli space of collective coordinates is
obtained from computing the inner products of the zero modes \cite{B},
see also \cite{BVvN} for a review. Semiclassically, the integration over
the quantum fluctuations yields one-loop determinants that have to be
evaluated in the instanton background. Finding this measure is an
essential step for computing correlation functions nonperturbatively.

How much of these field theory techniques can be applied to NS5-brane
instantons? First, we should keep in mind that we are approximating
the wrapped fivebrane by a four-dimensional supergravity instanton
solution. The (bosonic) collective coordinates are then just the
positions of the instanton in four-dimensional Euclidean space. In
string theory, the NS5-brane instanton is described by an embedding of
the worldvolume into the ten-dimensional product space $\fieldR^4\times
\mathrm{CY}$, such that the worldvolume of the fivebrane is wrapped
around the entire CY\@. The maps describing the embedding are thought
of as the collective coordinates of the instanton. Integrating over the
moduli space would then involve doing a path-integral over the
worldvolume theory in the supergravity background $\fieldR^4\times
\mathrm{CY}$. This is the general strategy advocated in \cite{BBS}.
This procedure is cumbersome, however, due to the complicated nature
of the NS-fivebrane worldvolume theory. In this paper, we have not
included any such worldvolume effects; we describe here the moduli
space as the finite dimensional space with coordinates $x_0^\mu$ and
their fermionic partners $\xi$ and $\bxi$ as discussed in the previous
section.

Second, there are the one-loop determinants. In our approach, these
determinants should be computed in the supergravity theory. They receive
corrections not only from the hypermultiplet fluctuations, but also from
the fluctuations of the gravitational and $h_{1,1}$ vector multiplets.
Their classical values in the instanton background are trivial (flat
metric, and vanishing vector multiplets), but their quantum effects
cannot be ignored. This is a complicated calculation which lies beyond
the scope of this paper. Moreover, these (and higher-) loop effects
would have to be computed in the full ten-dimensional string theory;
it remains to be seen if such a calculation can be done. As we indicate
below, we shall simply denote the determinants by $K$ and leave them
unspecified in further calculations. Clearly a better understanding of
instanton calculations in string theory is needed. We have taken here a
more pragmatic approach, and follow a line of thinking somewhat similar
to what Salviati and Sagredo conclude from their discussion in
\cite{HM2}.

We now proceed to calculate the measure. The zero modes are, by
definition, the zero eigenvalue eigenfunctions of the operator
sandwiched between the one-loop quantum fluctuations, and can be
obtained by taking the derivative of the instanton solution with respect
to the collective coordinates. We will here carry out the procedure of
computing the metric on the moduli space of single-centered instantons.
The case of anti-instantons is completely analogous.

There are two modifications with respect to the calculation in
\cite{B}. First, we are dealing with a non-linear sigma model in the
scalar field sector, where nontrivial metrics $\cG_{AB}$ and $M^{IJ}$
appear. Second, we are working with 2-form tensor fields which have to
be properly gauge-fixed, and whose zero modes will need to satisfy a
corresponding background gauge condition. In the first part of this
section we repeat the analysis of \cite{B}, applied to our system, and
in the second part we deal with the fermionic sector.

\bigskip
\textbf{Bosonic measure} 
\medskip

If we denote the bosonic fields of the double-tensor multiplet
collectively by
 \begin{equation}
  \Phi^M = \{ \phi^A, B_{\mu\nu I} \}\ ,
 \end{equation}
and expand the action about the instanton solution to quadratic order
in the fluctuations
 \begin{equation}
  \Phi^M = \Phi^M_\cl + \Phi^M_\mathrm{qu}\ ,
 \end{equation}
we can write (with flat background metric $g_{\mu\nu}=\delta_{\mu\nu}$)
 \begin{equation}
  S = S_\cl + \frac{1}{2}\, \int\! \d^4x\ \Phi^M_\mathrm{qu}\ \cG_{MP}
  \, \Delta^{P}{}_N\, \Phi^N_\mathrm{qu} + O(\Phi_\mathrm{qu}^3)\ .
 \end{equation}
Here, we have denoted (suppressing spacetime indices in the tensor
sector)
 \begin{equation} \label{G_MN}
  \cG_{MN} = \lp \cG_{AB}(\phi_\cl) & 0 \\[2pt] 0 & M^{IJ} (\phi_\cl)
  \rp\ .
 \end{equation}
$\Delta^M{}_N$ is a hermitean operator with respect to the inner product
defined by $\cG_{MN}$, and can be determined by explicitly expanding in
the fluctuations. For the moment it suffices to say that it takes the
form (again suppressing spacetime indices)
 \begin{equation} \label{Delta}
  \Delta^M{}_N = \lp -\delta^A{}_B\, \p^2 + \dots & * \\[2pt] * &
  -\delta_I{}^J\, \partial ^2 + \dots \rp\ ,
 \end{equation}
where the ellipsis stands for terms with operators at most linear in
derivatives. The off-diagonal terms are not written explicitly, but can
also be seen to be at most linear in derivatives.

Clearly, this operator is some sort of generalized Laplace operator, and
we assume it has a basis of orthogonal eigenfunctions $F^M_i$ in which we
expand the fluctuations,
 \begin{equation}
  \Delta^M{}_N F^N_i = \ep_i F^M_i\ ,\qquad \Phi^M_\mathrm{qu} = \sum_i
  \xi_i F^M_i\ .
 \end{equation}
The zero modes $F_{i_0}^M$ are eigenfunctions with zero eigenvalues
$\ep_{i_0}=0$, but nonzero fluctuation coefficients $\xi_{i_0}$. The
norms of the eigenfunctions are taken with respect to the metric
$\cG_{MN}$,
 \begin{equation}
  U_{ij} \equiv \langle F_i\,|F_j \rangle \equiv \int\! \d^4x\, F^M_i
  \cG_{MN} F^N_j\ .
 \end{equation}
The metric on the moduli space is found by computing the inner product
of the zero modes, and the latter can be obtained by taking derivatives
of the instanton solution with respect to the collective
coordinates\footnote{While such a derivative always gives a zero mode
(modulo gauge transformations, see below), it is to our knowledge
unclear whether in general there could be more zero modes than
collective coordinates.}. In the case of the single-centered instanton,
these are just the positions $x_0^\mu$ of the instanton in $\fieldR^4$.
Thus, there are four zero modes and the moduli space metric $(U_0)_{\mu
\nu}$ is four-dimensional, with contributions from both the scalars and
tensors. Since $\cG_{MN}$ is block-diagonal, there is no mixing between
the zero modes of these two sectors.

In the scalar sector the metric $\cG_{AB}$ is diagonal, so we can
consider the contributions from $\phi$ and $\chi$ separately. For the
dilaton zero mode $\p\phi/\p x_0^\mu=\e^\phi\p_\mu\e^{-\phi}$ we find,
using the spherical symmetry of the single-centered instanton,
 \begin{equation}
  U_{\mu\nu}^{(\phi)} = \int\! \d^4x\, \frac{\p\phi}{\p x_0^\mu}\,
  \frac{\p\phi}{\p x_0^\nu} = \int\! \d^4x\, \frac{x_\mu x_\nu}{r^2}\,
  \e^{2\phi} (\p_r \e^{-\phi})^2 = \frac{1}{4}\, \de_{\mu\nu} \int\!
  \d^4x\, \e^{2\phi} (\p_r \e^{-\phi})^2\ .
 \end{equation}
The integral formula \eqref{I_p} now yields the result
 \begin{equation}
  U_{\mu\nu}^{(\phi)} = \frac{|Q_2|}{4\, g_s^2}\, \de_{\mu\nu}\ .
 \end{equation}
Analogously, from \eqref{chi_constr} it follows for the $\chi$ zero mode
$\p\chi/\p x_0^\mu=g_s^2\Delta\chi\,\e^{2\phi}\p_\mu\e^{-\phi}$ that
 \begin{equation}
  U_{\mu\nu}^{(\chi)} = \int\! \d^4x\, \e^{-\phi} \frac{\p\chi}{\p
  x_0^\mu}\, \frac{\p\chi}{\p x_0^\nu}  = \frac{|Q_2|}{8}\, (\Delta
  \chi)^2\, \de_{\mu\nu}\ .
 \end{equation}
Here the $\e^{-\phi}$ insertion is the $\cG_{\chi\chi}$-component of the
metric \eqref{G_MN}.

The complete moduli space metric is the sum of the above $U$s and those
of the tensors, which we shall compute now. Being subject to gauge
symmetries, we first have to gauge-fix the tensors. We shall impose the
background gauge condition
 \begin{equation} \label{Bgf}
  \p^\mu \big( M^{IJ} B_{\mu\nu J}^\mathrm{qu} \big) = 0\ .
 \end{equation}
This requires a corresponding modification of $\Delta^M{}_N$ and the
inclusion of ghosts. The instanton configurations are solutions to the
classical, gauge-invariant equations of motion only, so derivatives with
respect to the collective coordinates in general do not yield zero modes
of the gauge-fixed operator $\Delta^M{}_N$. We need to add a suitable
gauge transformation to keep them in the background gauge,
 \begin{equation}
  Z_{\mu\nu I\rho} = \frac{\p B_{\mu\nu I}}{\p x_0^\rho} - 2\, \p_{[\mu}
  \Lambda_{\nu]I\rho}\ .
 \end{equation}
With $\Lambda_{\nu I\rho}=B_{\nu\rho I}$, we obtain
 \begin{equation}
  Z_{\mu\nu I\rho} = -H_{\mu\nu\rho I} = \ep_{\mu\nu\rho\si} H^\si_I\ .
 \end{equation}
These $Z_{\mu\nu I\rho}$ satisfy the gauge condition \eqref{Bgf} by
virtue of the classical tensor field equations. It follows that they
are zero modes of the gauge-fixed $\Delta^M{}_N$. Note that we do not
have to solve explicitly for the gauge potentials to compute their
zero mode norms, knowing the field strengths is sufficient. We now
calculate
 \begin{equation}
  U_{\mu\nu}^{(B)} = \frac{1}{2} \int\! \d^4x\, M^{IJ} Z^{\rho\si}\!
  {}_{I\mu}\, Z_{\rho\si J\nu} = \int\! \d^4x\, M^{IJ} (\de_{\mu\nu}
  H^\rho_I H_{\rho J} - H_{\mu I} H_{\nu J})\ .
 \end{equation}
Using spherical symmetry and the Bogomol'nyi condition \eqref{O2BPS},
we obtain
 \begin{equation}
  U_{\mu\nu}^{(B)} = \frac{3}{4}\, \de_{\mu\nu} \int\! \d^4x\,
  (\e^{\phi} \hat{H}^\rho_1 \hat{H}_{\rho 1} + \e^{2\phi} H^\rho_2
  H_{\rho 2}) = 3 \big( U_{\mu\nu}^{(\chi)} + U_{\mu\nu}^{(\phi)}
  \big)\ .
 \end{equation}
The sum of the scalar and tensor parts finally gives
 \begin{equation}
  (U_0)_{\mu\nu} = U_{\mu\nu}^{(\phi)} + U_{\mu\nu}^{(\chi)} +
  U_{\mu\nu}^{(B)} = S_\cl\, \de_{\mu\nu}\ ,
 \end{equation}
with $S_\cl$ as in \eqref{one-inst-act}, a result also familiar from
Yang-Mills instantons!

The bosonic part of the single-centered (anti-) instanton moduli space
measure is therefore
 \begin{equation}\label{bos-meas}
  \int\frac{\d^4x_0}{(2\pi)^2}\ (\det U_0)^{1/2}\, \e^{-S^\pm_\inst}
  (\det{\!}' \Delta)^{-1/2} = \int\frac{\d^4x_0}{(2\pi)^2}\
  S_\cl^2\, \e^{-S^\pm_\inst} (\det{\!}' \Delta)^{-1/2}\ ,
 \end{equation}
where $\det{\!}'\Delta$ stands for the amputated determinant, the
product of all nonzero eigenvalues. As explained in the beginning of
this section, computing this determinant lies beyond the scope of this
paper.

\bigskip
\textbf{Fermionic measure} 
\medskip

We now expand the fermion terms in the action about the (bosonic)
instanton solution, up to quadratic order in the fluctuations. Similarly
to the bosonic case, it is sufficient to consider first fluctuations in
the matter fermions $\la^a$ only, and freeze the fluctuations from
other multiplets, such as those coming from the gravitinos $\psi_\mu^i$.
We can then write
 \begin{equation} \label{quadr-ferm}
  S_2 = \int\! \d^4x\ \I\, \la^a_\mathrm{qu} (\slsh{D}_{3/4})_{a\bb}\,
  \bla^\bb_\mathrm{qu}
 \end{equation}
for the quadratic part of the action, where
 \begin{equation}
  (\slsh{D}_{3/4})_{a\bb} = \lp \slsh{D}_{3/4}\ & 0 \\[2pt] 0\ &
  \slsh{D}_{-3/4} \rp\ ,
 \end{equation}
and $\slsh{D}_k$ is defined in \eqref{Dk}. We have shown in sections
\ref{sect_DTM} and \ref{sect_FZM} that \eqref{quadr-ferm} indeed
produces the correct field equations.

We can construct hermitean operators $M_k=\slsh{D}_{\!-k}\,\slsh{\bar
D}_k$ and ${\bar M}_k=\slsh{\bar D}_{\!-k}\,\slsh{D}_k$. The cases of
interest are when $k=\pm 3/4$. The spectrum of nonzero modes of
$M_{-3/4}$ and ${\bar M}_{3/4}$ is identical, and similarly for
$M_{3/4}$ and ${\bar M}_{-3/4}$. This can be seen as follows:
Let $F^1_i$ and $F^2_i$ denote a basis of eigenfunctions of $M_{-3/4}$
and $M_{3/4}$ respectively, and ${\bar F}^1_i$ and ${\bar F}^2_i$ a
basis of eigenfunctions of ${\bar M}_{3/4}$ and ${\bar M}_{-3/4}$
respectively. The eigenfunctions of $M_{-3/4}$ and ${\bar M}_{3/4}$
are then related, with the same eigenvalue $\ep_i^1=\bar\ep_i^1\neq 0$,
by ${\bar F}^1_i=(\ep_i^1)^{-1/2}\slsh{\bar D}_{\!-3/4}F^1_i$, or
inversely, $F^1_i=(\ep_i^1)^{-1/2}\slsh{D}_{3/4}{\bar F}^1_i$.
Similarly, the spectrum of nonzero modes of $M_{3/4}$ and ${\bar
M}_{-3/4}$ is identical, and the relation between the eigenfunctions is
given by ${\bar F}^2_i=(\ep_i^2)^{-1/2}\slsh{\bar D}_{3/4}F^2_i$, with
inverse $F^2_i=(\ep_i^2)^{-1/2}\slsh{D}_{\!-3/4}{\bar F}^2_i$. We here
assumed for simplicity that the eigenvalues are positive, the argument
is similar for negative eigenvalues. Bearing in mind that both $M_{3/4}$
and ${\bar M}_{3/4}$ have zero modes, together with the fact that the
fermion zero modes are in $\la^2$ and $\bla^1$, we can expand the
fermions in a basis of eigenfunctions (suppressing spinor indices),
 \begin{equation}
  \la^a_\mathrm{qu} = \sum_i\, \xi^a_i F_i^a\ ,\qquad
  \bla^\ab_\mathrm{qu} = \sum_i\, \bar{\xi}_i^a \bar{F}^\ab_i\ ,
 \end{equation}
with $\xi_i^a$ and $\bar{\xi}^a_i$ anticommuting (there is no sum over
$a$). Plugging this into the action, and using the relation between the
different eigenfunctions as discussed above, we get
 \begin{equation}
  S_2 = \I \sum_{a,i,j} \xi^a_i\, U_{ij}^{aa}\, (\ep_j^a)^{1/2}\,
  \bar{\xi}^a_j\ ,\qquad U_{ij}^{ab} \equiv \int\! \d^4x\ F_i^aF_j^b\ .
 \end{equation}
We then define the fermionic part of the path-integral measure as
(up to a sign from the ordering of the differentials)
 \begin{equation}
  [\d\la]\, [\d\bla] \equiv \prod_a \prod_i\, \d\xi_i^a\, \d
  \bar{\xi}_i^a\, (\det U^{aa})^{-1}\ ,
 \end{equation}
such that the fermion integral gives the Pfaffians of $\slsh{\bar
D}_{3/4}$ and $\slsh{D}_{3/4}$ in the nonzero mode sector. In the zero
mode sector, we are left over with an integral over the four GCC, which
are combined into two spinors, multiplied by the inverses of the norms
of the zero modes. These zero mode eigenfunctions have the form
$Z^2_{\alpha\beta'}=\p{}^{(1)}\!\la^2_\alpha/\p\bxi^{\beta'}$ given in
\eqref{O2ZM_xi}, so that we find for their inner product
 \begin{align}
  U^{22}_{\alpha'\beta'} & = \int\! \d^4x\, {Z^{2\ga}}_{\alpha'}\,
    Z^2_{\ga\beta'} = -4 \int\! \d^4x\, \e^{-\phi/2} \p_\mu \phi\,
    \p_\nu \phi\, (\ep\,\bsi^\mu \si^\nu)_{\alpha'\beta'} \notag \\
  & = 4\, \ep_{\alpha'\beta'} \int\! \d^4x\, \e^{3\phi/2} (\p_r
    \e^{-\phi})^2 = \frac{8\, |Q_2|}{g_s}\, \ep_{\alpha'\beta'}\ .
 \end{align}
The fermionic measure on the moduli space of collective coordinates
then is
 \begin{equation} \label{ferm-meas}
  \int\! \d^2\xi\, \d^2\bar{\xi}\ \Big(\frac{g_s}{8\,|Q_2|}\Big)^2\
  \big( \det{\!}' M_{3/4}\ \det{\!}' \bar{M}_{3/4} \big)^{1/2}\ .
 \end{equation}
Here, our convention is that $\d^2\xi\equiv\d\xi_1\,\d\xi_2$.

The complete measure is then given by \eqref{ferm-meas} combined with
\eqref{bos-meas}, and has still to be supplemented with the gravitino,
supersymmetry ghost, and other fermionic determinants from the vector
multiplets. In the following we denote with $K_\text{1-loop}^\pm$ the
ratio of all fermionic and bosonic determinants in the
one-(anti-)instanton background. The single-centered (anti-) instanton
measure is then
 \begin{equation} \label{meas}
  \int \frac{\d^4x_0}{(2\pi)^2}\ \int\! \d^2\xi\, \d^2\bxi\ \Big(
  \frac{g_s\, S_\cl}{8\,|Q_2|} \Big)^2\, K_\text{1-loop}^\pm\,
  \e^{-S^\pm_\inst}\ .
 \end{equation}
This measure is the starting point for computing instanton corrections
to certain correlators. All the preparation is now done, and we can
finally focus on the explicit calculation of correlation functions.

\section{Correlation functions} \label{sect_correl}

We have found in  previous sections that there are four Grassmann
collective coordinates (GCC), and that they are all associated to the
broken supersymmetries. Also, the path integral measure contains an
integration over all collective coordinates, including the GCC. Hence,
a generic correlation function will only be non-zero if the operators
inserted in the path integral saturate the GCCs of the measure. It is
then clear that there will be a non-zero four-point fermion correlation
function. Diagrammatically, such a four-point vertex consists of four
fermion zero modes connected to an instanton at position $x_0$ which is
integrated over. Computing this diagram, one could read off the 4-index
tensor that determines the four-fermi terms in the effective action. As
explained in section \ref{sect_rigid}, this procedure is a bit
complicated due to the fact that we are working in a 1.5 order
formalism, where additional four-fermi terms are hidden in the spacetime
curvature scalar $R(\omega)$ as a function of the spin connection.
Moreover, the four-fermi correlator would merely determine (target
space) curvature-like terms, rather than the objects $M^{IJ}$, $\cG_{A
B}$ and $A^I_A$ which we are really interested in. Luckily, there is a
way out of this by studying the GCC dependence of the scalars and
tensors.

\bigskip
\textbf{GCC dependence of scalars and tensors} 
\medskip

We can in fact compute the instanton corrections to $M^{IJ}$, $\cG_{AB}$
and $A^I_A$ more directly by using again the broken supersymmetries.
These generate fluctuations which, by supersymmetry, are related to the
purely bosonic instanton, and are genuine zero modes which leave the
instanton action unchanged. The infinitesimal broken supersymmetries at
linear order induce the fermionic zero modes discussed above. The full
broken supersymmetry group is found by exponentiating the infinitesimal
transformations \eqref{susy-dtm}, and acting with them on the bosonic
instanton, we induce a GCC dependence of the scalars and tensors.
Expanding to second order, there will be a quadratic GCC dependence, and
this is sufficient for our purpose. The relevant correlators then will
be 2- and 3-point functions of scalars and tensors. Diagrammatically,
they correspond to instanton corrected propagators. These diagrams are
related by supersymmetry to the above mentioned four-fermi vertex.

At second order in the GCC, the scalars are given by
 \begin{equation}
  {}^{(2)}\!\phi^A = \frac{1}{2}\, \de_\eps^2 \phi^A|_\cl = \frac{1}{\2}
  \big( \ga^A_{ia}(\phi_\cl)\, \eps^i\, {}^{(1)}\!\la^a + \bar{\ga}^i
  {}^A_\ab(\phi_\cl)\, \beps_i\, {}^{(1)}\!\bla^\ab \big)\ .
 \end{equation}
The numerical factors come from the fact that the second supersymmetry
variation comes with a factor of 1/2, and from the rescaling of $\eps$
by a factor of $\sqrt 2$, cf.\ footnote \ref{ftnote3}. Using the
notation introduced at the end of section \ref{sect_FZM}, we find that,
since ${}^{(1)}\!\la^\pm={}^{(1)}\!\bla^\mp=0$ and for the broken
supersymmetries $\eps^\pm=\beps_\mp=0$, only terms proportional to
$\ga^A_{\mp\mp}$ and $\bar{\ga}^\pm{}_\pm^A$ contribute. For the dilaton
these are both zero, so only $\chi$ gets corrected at this order:
 \begin{equation}\label{2chi}
  {}^{(2)}\!\chi = 2\I\, \p_\mu \phi\, \xi \si^\mu \bxi\ ,\qquad
  {}^{(2)}\!\phi = 0\ .
 \end{equation}
Due to our conventions for the fermionic zero modes chosen in
\eqref{summ-ferm-zeromodes}, this expression for $\chi$ is the same
in the instanton and anti-instanton background.

Analogously, the second order corrections of the tensors follow from
\eqref{de_B_loc}. The instanton and anti-instanton cases yield, up to
a sign, the same answer,
 \begin{equation} \label{2B1}
  {}^{(2)}\!B_{\mu\nu1} = \mp 2\I\, \ep_{\mu\nu\rho\si}\, \p^\rho
  \e^{-\phi}\, \xi \si^\si \bxi\ ,\qquad {}^{(2)}\!B_{\mu\nu2} = 0\ .
 \end{equation}
Notice again that only the RR sector is turned on.

It turns out that the Bogomol'nyi equation \eqref{O2BPS} still holds
at this order in the GCC\@, for one can easily check that
 \begin{equation}
  {}^{(2)}\!H_{\mu1} = \pm \p_\mu \big( \e^{-\phi}\, {}^{(2)}\!\chi
  \big)\ .
 \end{equation}
The second component of \eqref{O2BPS} is trivially satisfied. It might
surprise the reader, who is somewhat familiar with instanton calculus,
that the equations of motion are satisfied without any fermion-bilinear
source term. One would expect such a source term to be present, since
\eqref{2chi} and \eqref{2B1} are obtained by acting with the broken
supersymmetries that also generate the fermionic zero modes. This is
typically what happens with the Yukawa terms in $N=2$ or $N=4$ SYM
theory in flat space; there the adjoint scalar field is found from
solving the inhomogeneous Laplace equation with a fermion-bilinear
source term. The fermionic zero modes in the presence of a YM instanton
then determine the profile and GCC dependence of the adjoint scalar
field. Some references where this is discussed in more detail are given
in \cite{DHKM}.

In the case at hand, the fermion bilinear source term actually vanishes
when the zero modes are plugged in. To see this, let us first consider
the tensors, for which the full equations of motion read
 \begin{align}
  e^{-1}\, \frac{\de S}{\de B_{\mu\nu I}} = \ep^{\mu\nu\rho\si} \p_\rho
    \big[ & M^{IJ} \cH_{\si J} - \I A_A^I \p_\si \phi^A +
    \frac{\I}{\2} M^{IJ} (g_{Jia} \psi_\si^i \la^a + \text{c.c.})
    \notag \\*
  & + \I M^{IJ} k_{Ja\ab}\, \la^a \si_\si \bla^\ab\, \big]\ .
 \end{align}
The fermionic zero modes we have found above do not enter these
equations directly, because ${}^{(1)}\psi_\mu^i={}^{(1)}\bpsi_{\mu
i}=0$, and the two matrices $M^{IJ}k_{Ja\ab}$ are diagonal (actually
zero for $I=1$), but for $a=\ab$ either ${}^{(1)}\!\la^a$ or ${}^{(1)}
\!\bla^\ab$ vanishes. Hence, up to second order in the GCC, only the
bosonic fields contribute. This is consistent with the fact that the
BPS condition still holds at this order. A similar analysis can be
done for the equations of motion for the scalars.

\bigskip
\textbf{2-point functions} 
\medskip

The relevant objects for computing correlation functions are products of
operators which saturate the GCC integral. These can be bilinears in the
RR fields ${}^{(2)}\!\chi$ and ${}^{(2)}\!H^\mu_1$, a combination of one
RR field with the zero modes ${}^{(1)}\la^\mp$ and ${}^{(1)}\bla^\pm$,
or the product of all four zero modes. Moreover, an arbitrary number of
background fields may be inserted. As explained in the beginning of this
section, we shall not compute the fermionic 4-point function, since we
can learn much more from the 2- and 3-point functions.

The first step is to take the large distance limit of the zero modes and
express them in terms of propagators, which will enable us to read off
the effective vertices from the correlation functions by stripping off
the external legs. For the bosons we find
 \begin{align}
  {}^{(2)}\!\chi(x) & = - 2\I\, |Q_2|\, g_s^{-2}\, \xi \si^\mu \bxi\,
    \p_\mu G(x,x_0)\, \big( 1 + \dots \big) \notag \\[2pt]
  {}^{(2)}\!H^\mu_1(x) & = \mp 2\I\, |Q_2|\, \xi \si^\nu \bxi\, \big(
    \p^\mu \p_\nu - \de^\mu_\nu \p^2 \big) G(x,x_0)\ , \label{eq:exact1}
 \end{align}
where $G(x,x_0)=1/4\pi^2(x-x_0)^2$ is the massless scalar propagator.
The second equation is exact, while from the first we only keep the
leading term in the large distance expansion valid when $(x-x_0)^2\gg
|Q_2|/4\pi^2 g_s^2$. In this limit the dilaton is effectively given by
$\e^{-\phi}\approx\e^{-\phi_\infty}=g_s^2$, and similarly $\chi\approx
\chi_\infty$, so the fields are replaced by their asymptotic values,
and these will be used to describe the asymptotic geometry of the moduli
space in the next section.

For the fermions we have (the ellipsis again indicates terms beyond
the large distance expansion)
 \begin{align} \label{ferm_prop}
  {}^{(1)}\la^\mp_\alpha(x) & = - 2\, |Q_2|\, g_s^{-3/2} S_{\alpha
    \beta'} (x,x_0)\, \bxi^{\beta'} \big( 1 + \dots \big) \notag
    \\[2pt]
  {}^{(1)} \bla^\pm_{\beta'}(x) & = \pm 2\, |Q_2|\, g_s^{-3/2}\,
    \xi^\alpha S_{\alpha\beta'}(x,x_0)\, \big( 1 + \dots \big)\ ,
 \end{align}
where $S(x,x_0)=-\I{\p\!\!\!/}\,G(x,x_0)$ is the $\la\bla$ propagator.
The signs in the second equation reflect our choice of conventions for
instantons and anti-instantons, see \eqref{summ-ferm-zeromodes}.

Let us begin with the purely bosonic correlators: With the GCC measure
$\d\mu_\xi\equiv\d^2\xi\,\d^2\bxi$ $\big(g_s/8|Q_2|\big)^2$ from
\eqref{ferm-meas} and the Fierz identity $\xi\si_\mu\bxi\,\xi\si_\nu\bxi
=-\half\de_{\mu\nu}\,\xi\xi\,\bxi\bxi$, we find in the large distance
limit
 \begin{equation}
  \int\! \d\mu_\xi\, {}^{(2)}\chi(x)\, {}^{(2)}\chi(y) = \frac{1}{8
  g_s^2}\, \p_0^\mu\, G(x,x_0)\, \p^0_\mu\, G(y,x_0)\ ,
 \end{equation}
where $\p^0_\mu\equiv\p/\p x^\mu_0$ denotes the derivative with respect
to the bosonic collective coordinates. Using ${}^{(2)}\phi=0$, we then
obtain for the leading semiclassical contribution to the correlation
function of two scalars in the one-(anti-)instanton background
 \begin{align}
  \langle \phi^A(x)\, \phi^B(y) \rangle_\pm & = g_s^{-2}\, \de^A_\chi
    \de^B_\chi\! \int\! \d^4x_0\ Y_\pm\, \p_0^\mu\, G(x,x_0)\,
    \p^0_\mu\, G(y,x_0) \notag \\[2pt]
  & = g_s^{-2}\, Y_\pm\, \de^A_\chi \de^B_\chi\, G(x,y)\ .
    \label{cor_chi-chi}
 \end{align}
Here we denote (remember that the difference between $S_\cl$ and
$S^\pm_\inst$ is given by the surface terms \eqref{theta-angles2})
 \begin{equation} \label{Y-inst}
  Y_\pm \equiv \frac{1}{32\pi^2}\, S_\cl^2\, \e^{-S^\pm_\inst}
  K_\text{1-loop}^\pm\ ,
 \end{equation}
which is small for small string coupling constant $g_s$. Since
translation invariance implies that neither $S_\cl$ nor
$K_\text{1-loop}^\pm$ depend on the collective coordinates $x_0$, we
were allowed to integrate by parts and use $\p_0^2\,G(x,x_0)=-\de(x-
x_0)$. There is no boundary term because the domain of integration
covers all of $\fieldR^4$ with no points excised (it is an integral
over moduli space, not spacetime), and the integrand vanishes at
infinity. The result \eqref{cor_chi-chi} is to be compared with the
propagator derived from an effective action with instanton and
anti-instanton corrected metric $\cG_{AB}^\eff=\cG_{AB}+\cG_{AB}^\inst$,
with $\cG_{AB}$ as in \eqref{DTM_metrics}. Similarly we write for the
inverse $\cG^{AB}_\eff=\cG^{AB}+\cG^{AB}_\inst$, with $\cG_{AC}\cG^{CB}
=\delta_A^B$. At leading order, we find
 \begin{equation} \label{inv-cG_eff}
  \cG^{AB}_\inst = \lp 0 & 0 \\[2pt] 0 & g_s^{-2} (Y_+ + Y_-) \rp\ .
 \end{equation}
Note that since $Y_-=(Y_+)^*$, instanton and anti-instanton
contributions combine into a real correction\footnote{We are assuming
here that $K^-=(K^+)^*$. Presumably, the one-loop determinants $K^\pm$
only differ by a phase coming from the fermionic determinants. If this
phase can be absorbed in the corresponding surface terms
\eqref{theta-angles2}, the instanton and anti-instanton determinants are
real and equal. \label{ftnote10}}. This result receives of course
corrections from perturbation theory and from terms that become
important beyond the large distance approximation in which $\e^{-\phi}
\approx g_s^2$. Such terms play a role when inverting the result of
\eqref{inv-cG_eff} to obtain the effective metric $\cG_{AB}^\eff$. They
correspond to higher order powers in $Y_\pm$ and interfere with
multi-centered (anti-) instanton effects. Dropping all these subleading
terms, which is the approximation which we are working in, we find
 \begin{equation} \label{cG_eff}
  \cG_{AB}^\eff =  \lp 1 & 0 \\[2pt] 0 & \e^{-\phi} - g_s^2 (Y_+ + Y_-)
  \rp \ .
 \end{equation}

For two RR tensors we find similarly
 \begin{equation}
  \int\! \d\mu_\xi\, {}^{(2)}H_{\mu1}(x)\, {}^{(2)}H_{\nu1}(y) =
  \frac{g_s^2}{8}\, G_{\mu\rho}(x,x_0)\, G^\rho_\nu(y,x_0)\ ,
 \end{equation}
where $G_{\mu\nu}(x,x_0)=\big(\p_\mu\p_\nu-\de_{\mu\nu}\p^2\big)G(x,
x_0)$ is the gauge-invariant propagator of dual tensor field strengths.
Using ${}^{(2)}H_{\mu2}=0$ and the convolution property $G_{\mu\rho}*
G_\nu^\rho=G_{\mu\nu}$, it follows that
 \begin{equation}
  \langle H_{\mu I}(x)\, H_{\nu J}(y) \rangle_\pm = g_s^2\, Y_\pm\,
  \de_I^1\, \de_J^1\, G_{\mu\nu}(x,y)\ .
 \end{equation}
{}From the right-hand side we read off the (anti-) instanton correction
to the inverse metric $M_{IJ}$, which multiplies the tensor propagators.
We find for the sum
 \begin{equation} \label{M_eff}
  M_{IJ}^\inst = g_s^2\, (Y_+ + Y_-)\, \de_I^1\, \de_J^1\ ,
 \end{equation}

In the large distance approximation, where also $\chi\approx
\chi_\infty$, we then obtain
 \begin{equation}
  M^{IJ}_\eff = M^{IJ}- g_s^{-2}\, (Y_+ + Y_-) \lp 1 & -\chi_\infty
  \\[2pt] -\chi_\infty & \chi^2_\infty \rp\ ,
 \end{equation}
with $M^{IJ}$ as in \eqref{DTM_metrics}. This naively suggests that both
RR and NS-NS sectors get corrections in front of the tensor kinetic
terms. However, when expressed in terms of $\hat{H}_1=H_1-\chi H_2$,
the tensor kinetic terms in the effective action simplify to
 \begin{equation}
  e^{-1} \cL_\eff = \frac{1}{2}\, \big( \e^\phi - g_s^{-2}
  (Y_+ + Y_-) \big)\, \hat{H}^\mu_1 \hat{H}_{\mu1} + \frac{1}{2}\,
  \e^{2\phi} H^\mu_2 H_{\mu2} + \dots\ ,
 \end{equation}
In this basis, which is the one to distinguish between fivebrane and
membrane instantons (see the discussion in section \ref{dtm-section}),
the NS-NS sector does not receive any instanton corrections.

Last but not least, the mixed bosonic combination
 \begin{equation} \label{H-chi}
  \int\! \d\mu_\xi\, {}^{(2)}H_{\mu1}(x)\, {}^{(2)}\chi(y) = \mp
  \frac{1}{8}\, G_{\mu\nu}(x,x_0)\, \p_0^\nu\, G(y,x_0)
 \end{equation}
obviously vanishes when integrated over $x_0$ thanks to the Bianchi
identity $\p^\mu G_{\mu\nu}=0$. We conclude that
 \begin{equation}
  \langle H_{\mu I}(x)\, \phi^A(y) \rangle = 0\ .
 \end{equation}
This was to be expected, since for constant $A_A^I$ the vertex $-\I
A_A^I H^\mu_I\p_\mu\phi^A$ is a total derivative and therefore does not
contribute to the propagator. However, in the next section we show that
instantons do induce such a vertex with field-dependent coefficients.


\bigskip
\textbf{3-point functions} 
\medskip

While we cannot determine the coefficients $A_A^I$ directly, we can
compute the field strength ${F_{AB}}^I=2\p_{[A}^{}A_{B]}^I$ from
suitable correlation functions and then integrate it. To lowest
nontrivial order in an expansion of $A_A^I$ in powers of $\phi^A$, the
vertex $-\I A_A^I H^\mu_I\p_\mu\phi^A$ induces a 3-point function
 \begin{equation}
  \langle \phi^A(x)\, \phi^B(y)\, H_{\mu I}(z) \rangle = \I M^\infty_{
  IJ}\, \cG^{AC}_\infty\, \cG^{BD}_\infty {F_{CD}}^J\! \int\! \d^4x_0\,
  G(x,x_0)\, \p^\nu_0 G(y,x_0)\, G_{\mu\nu}(z,x_0)\ .
 \end{equation}
The antisymmetric derivative of $A_A^I$ arises here by virtue of the two
possible contractions of the scalars and integrating $\p^\nu_0$ by
parts. ${F_{CD}}^J$ denotes the constant part of the full field
strength. The prefactors $M_{IJ}^\infty$ and $\cG^{AB}_\infty$ come from
the tensor and scalar propagators, respectively, where the sub- or
superscript $\infty$ indicates replacing the fields by their asymptotic
values at infinity. This 3-point function is to be compared with the
result of inserting the GCC-dependent fields. Due to the antisymmetry,
the two scalars must be different. Since $\phi$ has no GCC dependence,
they have to be contributed by $\chi$ and the tensor:
 \begin{equation}
  \langle \phi(x)\, \chi(y)\, H_{\mu I}(z) \rangle = \de_I^1\, \langle
  \phi_\cl(x)\, {}^{(2)}\chi(y)\, {}^{(2)}H_{\mu 1}(z) \rangle\ .
 \end{equation}
We expand $\phi_\cl$ as
 \begin{equation}
  \phi_\cl(x) = - 2 \ln(g_s) - \frac{|Q_2|}{g_s^2}\, G(x,x_0) + \dots\ .
 \end{equation}
The leading, constant term reduces the above to the 2-point function
\eqref{H-chi}, which vanishes. The next term in the expansion gives,
after integrating the GCC,
 \begin{equation}
  \langle \phi(x)\, \chi(y)\, H_{\mu I}(z) \rangle_\pm = \pm
  \frac{|Q_2|}{g_s^2}\, Y_\pm \de_I^1\! \int\! \d^4x_0\, G(x,x_0)\,
  \p^\nu_0 G(y,x_0)\, G_{\mu\nu}(z,x_0)\ .
 \end{equation}
We conclude that
 \begin{equation}
  \frac{|Q_2|}{g_s^2}\, (Y_+ - Y_-) \de_I^1 = \I M_{IJ}^\infty\,
  \cG^{\phi A}_\infty\, \cG^{\chi B}_\infty F_{AB}^{\inst J}\ ,
 \end{equation}
from which follows
 \begin{equation} \label{Fresult}
  F_{\phi\chi}^{\inst\,1} = -\I \frac{|Q_2|}{g_s^2}\, (Y_+ - Y_-)\ ,
  \qquad F_{\phi\chi}^{\inst\,2} = - \chi_\infty F_{\phi\chi}^{\inst\,
  1}\ .
 \end{equation}

Next, we compute instanton corrections to $\Gamma^{Ia}{}_b$. It appears
in the effective action \eqref{Eucl-rigid-action} through the relation
\eqref{k-Gamma}, and measures the strength of the coupling between the
tensors and the fermions. We therefore compute
 \begin{equation}
  \langle \la_\alpha^a(x)\, \bla_{\beta'}^\bb(y)\, H_{\mu I}(z)
  \rangle_\pm = -\I \frac{|Q_2|}{g_s}\, Y_\pm \de^a_\mp\, \de^\bb_\pm\,
  \de_I^1\! \int\! \d^4x_0\, \big[ S(x,x_0) \bsi^\nu S(y,x_0) \big]_{
  \alpha\beta'}\, G_{\mu\nu}(z,x_0)\ .
 \end{equation}
These two correlators induce an effective vertex $-h_{a\ab}
(\Gamma^I_\inst)^a{}_b\,\la^b\si^\mu\bla^\ab H_{\mu I}$ with
coefficients
 \begin{equation} \label{Gamma_inst}
  (\Gamma_\inst^I)^a{}_b = -\I \frac{|Q_2|}{g_s}\, M^{I1}_\infty\,
  (Y_+\, \de^a_2 h_{b1} + Y_-\, \de^a_1 h_{b2}) = -\I \frac{|Q_2|}
  {g_s}\, M^{I1}_\infty \lp 0 & Y_- \\[2pt] Y_+ & 0 \rp\ .
 \end{equation}
Here we have used that $h_{a\bb}$ is not corrected at leading
order\footnote{The 2-point function of two fermion insertions vanishes
in the semiclassical limit. Moreover, the result for the 3-point
function \eqref{lalaphi} shows that no field dependence of the metric
$h_{a\bb}$ is induced to leading order.}.

The last two correlators contribute to the connection $\G{A}{a}{b}$,
which appears in the covariant derivative on the fermions
\eqref{cov-der}. This connection was zero on tree-level (see
\eqref{DTM-connections}), but it receives instanton corrections as
follows from
 \begin{equation} \label{lalaphi}
  \langle \la_\alpha^a(x)\, \bla_{\beta'}^\bb(y)\, \phi^A(z) \rangle_\pm
  = \mp \I \frac{|Q_2|}{g_s^3}\, Y_\pm \de^a_\mp\, \de^\bb_\pm\,
  \de^A_\chi\! \int\! \d^4x_0\, \big[ S(x,x_0) \bsi^\mu S(y,x_0)
  \big]_{\alpha\beta'}\, \p_\mu\, G(z,x_0)\ .
 \end{equation}
This corresponds to an effective vertex $-\I h_{a\ab}(\Gamma^\inst_{\!A}
)^a{}_b\,\la^b\si^\mu\bla^\ab\,\p_\mu\phi^A$ with
 \begin{equation}
  (\Gamma^\inst_{\!A})^a{}_b = \frac{|Q_2|}{g_s^3}\, \cG_{A
  \chi}^\infty \lp 0 & -Y_- \\[2pt] Y_+ & 0 \rp\ .
 \end{equation}

The above connections induce instanton corrections to the curvature
tensors that appear in the four-fermi couplings of the effective action
\eqref{EuclDTMaction}. Indeed, that these curvatures receive instanton
corrections also follows from the computation of 4-point functions of
fermionic insertions, and these results should be consistent  with
computing the curvatures from the connections. For reasons explained in
the beginning of this section, checking this consistency may be a
complicated task.

Notice, however, that there is another four-fermi term in
\eqref{EuclDTMaction} proportional to the product of two antisymmetric
tensors $\cE_{ab}$. It is easy to see that this tensor cannot receive
instanton corrections, since it multiplies only $\la^a$ in the action,
not $\bla^\ab$. Due to the even distribution of fermionic zero modes
among $\la^a$ and $\bla^\ab$ there are thus no non-vanishing correlation
functions that could induce an effective vertex involving $\cE_{ab}$. A
similar argument shows that the connections $\Gamma^{Ii}{}_j$ do not get
corrected: they occur in the action only in combination with gravitinos
(e.g.\ in the vertex $2\Gamma^{Ii}{}_j H^{\mu\nu\rho}_I\psi^j_\mu
\si_\nu^{}\bpsi_{\rho i}$, hidden in the square of the supercovariant
field strengths of the tensors \cite{TV2}), which have no zero modes to
lowest order in the GCC\@. Correlation functions of fields corresponding
to vertices involving $\Gamma^{Ii}{}_j$ then do not saturate the GCC
integrals and vanish. If we were to continue the procedure of sweeping
out solutions by applying successive broken supersymmetry
transformations to the fields, the gravitinos may obtain a GCC
dependence at third order, but then the number of GCCs in the
correlators of interest exceeds the number of degrees of freedom and
they therefore vanish as well. Note, however, that due to
\eqref{Gamma-Omega} the coefficients ${\Omega_I}^i{}_j$ do get
corrected:
 \begin{equation}
  (\Omega_I^\eff)^i{}_j = M_{IJ}^\eff\, \Gamma^{Ji}{}_j =
  {\Omega_I}^i{}_j + g_s^2 (Y_+ + Y_-) \de_I^1\, \Gamma^{1i}_{\infty
  j}\ .
 \end{equation}
These quantities appear in the supersymmetry transformations of the
tensors \eqref{de_B_loc}.
\medskip

We can use the results for the various connections in an independent
derivation of the field strength components \eqref{Fresult} by means of
the identity \eqref{F},
 \begin{equation}
  F_{AB}{}^I = - 2 \Tr\big( W_{\!A}^\dag\, h^t\, \Gamma^I W_{\!B}^{}
  \big) - 2 \Tr\big( W_{\!A}^\dag\, h^t\, W_{\!B}^{} \Gamma^{It}
  \big)\ ,
 \end{equation}
which must hold also in the effective theory if supersymmetry is
preserved. The first trace on the right contains the connection
$\Gamma^{Ia}{}_b$, the second trace the connection $\Gamma^{Ii}{}_j$.
As argued above, the latter is not modified by instantons. Using the
corrections to the vielbeins derived in appendix \ref{inst-vielb}, it
is then readily verified that the second trace vanishes identically as
a result of $\Gamma^{Ii}{}_j$ being symmetric. On the other hand, the
first trace does receive a contribution from the instanton-corrected
connection \eqref{Gamma_inst}. A short calculation then yields the same
expressions as in \eqref{Fresult}.

Let us now determine the connections $A_A^I$. Toward this end, we use
invariance of the effective action under transformations $A_A^I
\rightarrow A_A^I+\p_A\xi^I(\phi)$ to choose a convenient gauge, namely
 \begin{equation} \label{Aphi}
  A_\phi^I = 0\ .
 \end{equation}
In this axial gauge we have ${F_{\phi\chi}}^I=\p_\phi A_\chi^I$.
Consider now the fields
 \begin{equation} \label{Achi}
  A_\chi^1 = \I\+ (Y_+ - Y_-)\ ,\qquad A_\chi^2 = - \chi_\infty
  A_\chi^1\ .
 \end{equation}
Using the relation $\p_{\phi_\infty}=g_s^{-2}\p/\p g_s^{-2}$, the
derivative with respect to the modulus $\phi_\infty$ of $Y_\pm$ given
in \eqref{Y-inst} yields
 \begin{align}
  32 \pi^2\, \p_{\phi_\infty} \I Y_\pm & = - \I \frac{|Q_2|}{g_s^2}\,
    S_\cl^2\, \e^{-S_\inst^\pm} K_\text{1-loop}^\pm + 2 \I
    \frac{|Q_2|}{g_s^2}\, S_\cl\, \e^{-S_\inst^\pm} K_\text{%
    1-loop}^\pm \notag \\
  & \tab + \I S_\cl^2\, \e^{-S_\inst^\pm}\, \p_{\phi_\infty} K_\text{%
    1-loop}^\pm\ .
 \end{align}
For small $g_s$, the second term on the right-hand side is suppressed
by a factor $g_s^2$ as compared to the first term. We can also assume
that the derivative of the 1-loop determinant gives only a subleading
contribution. In our approximation, the first term is the dominating one
and coincides exactly with the field strength \eqref{Fresult}. We
conclude that the expressions in \eqref{Aphi}, \eqref{Achi} are the
sought-after instanton-corrected connection coefficients.

\section{The universal hypermultiplet moduli space} \label{sect_modul}

In order to determine the instanton corrections to the universal
hypermultiplet, we first Wick-rotate back to Lorentz signature and then
dualize the tensors $H_I$ into two pseudoscalars $\phi^I=(\varphi,\si)$,
using the same notation as in the introduction, i.e.\ $\varphi$ is a RR
field and $\si$ the NS axion. If we combine the latter and $\phi^A=(\phi
,\chi)$ into a four-component field $\phi^{\hat{A}}=(\phi^A, \phi^I)$,
then in this basis the universal hypermultiplet metric reads \cite{TV2}
 \begin{equation}\label{dual-metric}
  G_{\!\hat{A}\hat{B}} = \lp \cG_{AB} + A_A^I M_{IJ} A_B^J & A_A^K
  M_{KJ} \\[2pt] M_{IK} A_B^K & M_{IJ} \rp\ .
 \end{equation}
Using \eqref{cG_eff}, \eqref{M_eff}, \eqref{Aphi} and \eqref{Achi}, we
find for the asymptotic effective Lagrangian
 \begin{align}
  e^{-1} \cL_\mathrm{UH} & = - \frac{1}{2}\, (\p_\mu \phi)^2 - \frac{1}
    {2}\, \e^{-\phi} (1 - g_s^2 \e^\phi Y)\, (\p_\mu \chi)^2 - \frac{1}
    {2}\, \e^{-\phi} (1 + g_s^2 \e^\phi Y)\, (\p_\mu \varphi)^2
    \notag \\*
  & \tab - \e^{-\phi} \tilde{Y}\+ \p_\mu \chi\, \p^\mu \varphi -
    \frac{1}{2}\, \e^{-2\phi} (\p_\mu \si + \chi \p_\mu \varphi)^2 +
    \dots\ , \label{lag1}
 \end{align}
where the ellipsis stands for subleading terms. $Y\equiv Y_++Y_-$ is the
sum of the instanton and anti-instanton contributions, as introduced in
\eqref{Y-inst}. It can be written as
 \begin{align} \label{total-Y}
  Y & = \frac{1}{32\pi^2}\, S_\cl^2\, \e^{-S_\cl} \big( \e^{-\I{\hat\si}
    |Q_2|} K_\text{1-loop}^+ + \e^{\I{\hat\si |Q_2|}} K_\text{1-loop}^-
    \big)\notag \\*[2pt]
   & = \frac{1}{16\pi^2}\, S_\cl^2\, \e^{-S_\cl} K_\text{1-loop}
    \cos(\hat\si Q_2)\ ,
 \end{align}
where we have introduced $\hat\si\equiv\si+\chi_0\varphi$ such that $Y$
is periodic in $\hat\si$. The second equality in \eqref{total-Y} holds
only under the assumption made in footnote \ref{ftnote10}. Similarly, we
have
 \begin{equation} \label{total-tildeY}
  \tilde{Y} \equiv \I\mskip1mu (Y_+ - Y_-) = \frac{1}{16\pi^2}\,
  S_\cl^2\, \e^{-S_\cl} K_\text{1-loop} \sin(\hat\si |Q_2|)\ .
 \end{equation}
To derive this term\footnote{We thank S.~Alexandrov for pointing
out a mistake in a previous version of this paper.}, we have used
that $\chi\approx\chi_\infty$, which holds in the large distance
approximation made in this paper. Notice furthermore that only the
RR sector receives corrections from the NS5-brane instanton.

\bigskip
\textbf{The metric and isometries} 
\medskip

The next step is to write down the line element, which is given by
 \begin{equation*}
  \d s_\mathrm{UH}^2 = G_{\!\hat{A}\hat{B}}\, \d\phi^{\hat{A}} \otimes
  \d\phi^{\hat{B}}\ .
 \end{equation*}
We remind the reader that the classical metric reads
 \begin{equation}
  \d s_\mathrm{UH}^2 = \d\phi^2 + \e^{-\phi} \d\chi^2 + \e^{-\phi}
  \d\varphi^2 + \e^{-2\phi} (\d\sigma + \chi \d\varphi)^2 \ ,
 \end{equation}
and describes the homogeneous quaternion-K\"ahler space $\SU(1,2)/
\U(2)$ \cite{CFG}. As mentioned in the introduction, the isometry group
$\SU(1,2)$ can be split into three categories. First, there is a
Heisenberg subgroup of shift isometries,
 \begin{equation} \label{Heis-alg2}
  \phi \rightarrow \phi \ ,\qquad \chi \rightarrow \chi + \gamma\ ,
  \qquad \varphi \rightarrow \varphi + \beta\ ,\qquad \sigma
  \rightarrow \sigma - \alpha - \gamma\, \varphi\ ,
 \end{equation}
where $\alpha$, $\beta$, $\gamma$ are real (finite) parameters. This
Heisenberg group is preserved in perturbation theory \cite{S}. We have
not discussed these corrections (which only appear at one-loop in the
string frame) here; they are discussed in \cite{AMTV,ARV} and should be
added to our final result for the metric.

Second, there is a U(1) symmetry that acts as a rotation on $\varphi$
and $\chi$, accompanied by a compensating transformation on $\sigma$.
Its finite transformation can be determined from the results in
\cite{BB,DDVTV} and reads
 \begin{gather}
  \varphi \rightarrow \cos\delta\,\, \varphi + \sin\delta\,\, \chi\ ,
  \qquad \chi \rightarrow \cos\delta\,\, \chi - \sin\delta\,\, \varphi
  \notag \\[2pt]
  \si \rightarrow \sigma - \tfrac{1}{4} \sin(2\delta)\,\, (\chi^2 -
  \varphi^2) + \sin^2\!\delta\,\, \chi \varphi\ . \label{delta-isom}
 \end{gather}
The remaining four isometries involve non-trivial transformations on
the dilaton, and hence will change the string coupling constant. Their
infinitesimal form was found in \cite{DWVP,DWVVP}, and we will not
explore the fate of these isometries nonperturbatively. In fact, at
the moment of writing this, it has not been analyzed whether these
dilaton-transforming isometries survive the perturbative corrections.

We now present the instanton corrected moduli space metric. As shown
above, instanton effects are proportional to $Y$ and $\tilde{Y}$ given
by \eqref{total-Y} and \eqref{total-tildeY}, and depend on the instanton
charge $Q_2$ and the RR background specified by $\chi_0$. Moreover, also
the asymptotic values of the fields, $g_s$ and $\chi_\infty$, appear;
they are treated as coordinates in the asymptotic regime of the moduli
space, i.e., where $\chi =\chi_\infty$ and $\e^{-\phi}=g_s^2$. For fixed
values of $\chi_0$ and $Q_2$, the moduli space metric is given by
 \begin{equation} \label{nonpert-metric}
  \d s_\mathrm{UH}^2 = \d\phi^2 + \e^{-\phi} (1-Y) \d{\chi}^2 +
  \e^{-\phi} (1+Y) \d\varphi^2 + 2\+ \e^{-\phi} \tilde{Y} \d\chi\+
  \d\varphi + \e^{-2\phi} (\d\si + \chi \d\varphi)^2\ ,
 \end{equation}
up to subleading terms. This metric therefore satisfies the constraints
from quaternionic geometry only up to leading order; to what extent the
quaternionic structure can fix these subleading corrections remains to
be seen. The result written in \eqref{nonpert-metric} depends on $Q_2$
and on the chosen RR background. To obtain the full moduli space metric,
one must sum over all instanton numbers $Q_2$. It would be very
interesting to do this sum explicitly, and to see of which function we
have the asymptotic limit. Unfortunately, for that we need more
knowledge on the one-loop determinants and the subleading corrections.

We can also deduce the leading-order instanton corrections to the
vielbeins and other geometric quantities. These can be computed from the
vielbeins that determine the double-tensor multiplet geometry, which we
give in appendix~\ref{inst-vielb}.

What happens to the isometries\footnote{We assume here that the
isometry transformations do not receive any quantum corrections.}
\eqref{Heis-alg2} and \eqref{delta-isom}? For the Heisenberg
group, this amounts to investigating which isometries are broken
by the quantities $Y$ and ${\tilde Y}$, as the other terms are
invariant. First we focus on the $\gamma$-shift in $\chi$. For a
given, fixed RR background $\chi_0$, the $\gamma$-shift is broken
completely. This is because $Y$ is proportional to $S_\cl$, which
contains $\Delta\chi=\chi_\infty-\chi_0$, see
\eqref{one-inst-act}. However, this symmetry can be restored if we
simultaneously change the background by $\chi_0\rightarrow\chi_0+
\gamma$. Since $\chi_0$ is subject to a quantization condition
(see section \ref{Fivebrane-inst}), this induces a quantization
condition on the possible values for $\gamma$. This means that the
$\gamma$-shift is broken to a discrete subgroup.

With this in mind, we find that under the action of a generic element
in the Heisenberg group the metric is invariant only if the following
quantization condition is satisfied:
 \begin{equation} \label{quant-cond}
  \alpha - (\chi_0 + \gamma) \beta = \frac{2\pi n}{|Q_2|}\ ,
 \end{equation}
with $n$ an integer. As explained before, the $\gamma$-dependence is
not relevant here since we could shift the RR background again. For the
other two isometries, generated by $\alpha$ and $\beta$, only a linear
combination is preserved. Stated differently, the $\beta$-isometry is
preserved as a continuous isometry if we accompany it by a compensating
$\alpha$-shift, where $\alpha$ is determined from \eqref{quant-cond}.

If we solely perform an $\alpha$-transformation, only a discrete
$\fieldZ_{|Q_2|}$ subgroup survives as a symmetry. In fact, since the
full metric includes a sum over $Q_2$, only shifts with $\alpha=
2\pi n$ are unbroken. In conclusion, for the Heisenberg group, one
isometry remains continuous, and two are broken to discrete subgroups.
This is precisely in line with the proposal made in \cite{ARV}.

The remaining isometry we discuss is \eqref{delta-isom}. Since the last
term in \eqref{nonpert-metric} is invariant by itself, we should only
look at the RR sector. Due to the fact that $Y$ is independent of
$\varphi$, but depends on $\chi^2$, this continuous rotation symmetry
seems to be broken. In fact, the terms proportional to $Y$ and
$\tilde{Y}$ break this isometry down to the identity $\delta=0$ and
the discrete transformation with $\delta=\pi$,
 \begin{equation}
  \chi \rightarrow -\chi\ ,\qquad \varphi \rightarrow -\varphi\ ,
  \qquad \si \rightarrow \si\ .
 \end{equation}
This conclusion is different from \cite{BB}, where also
$\delta=\pi/2$ was claimed to survive as an isometry. It is not
excluded, however, that in a full treatment the exact answer might
restore some of the broken symmetries. Clearly, this is an
interesting point that deserves further study.

Notice finally also the existence of another discrete isometry, which
changes the sign in $\chi$ (or $\varphi$) together with a sign flip
in $\si$. This is because the (leading) instanton plus anti-instanton
corrections are even in $\chi$ and $\si$. This discrete isometry is
however not part of (a discrete subgroup of) $\SU(1,2)$.

\section{Conclusions}

In this paper we have performed a detailed semiclassical computation of
certain correlators in the background of an NS5-brane instanton, in a
supergravity theory coupled to the universal hypermultiplet. We have
also studied the effects of turning on nontrivial RR background matter
fields. This has resulted in the instanton corrected moduli space for
the universal hypermultiplet metric in the asymptotic regime. Our main
result can be summarized by \eqref{nonpert-metric} together with
\eqref{total-Y}, and should be combined with the one-loop correction
found in \cite{AMTV}. An unexpected result is the fact that the presence
of the NS5-brane only affects the Ramond-Ramond sector. Our result for
the universal hypermultiplet metric has enabled us to discuss the
breaking of isometries, and in particular we have demonstrated the
breaking of the Heisenberg group to a discrete subgroup thereof, as
explained in the last section, and summarized by \eqref{quant-cond}.

Recently, a conjecture for the fivebrane instanton corrections to the
universal hypermultiplet moduli space metric was made in \cite{ARV},
using superspace techniques. It would be very interesting to test if
this proposal reproduces our results in the semiclassical limit. The
structure of the breaking of isometries is already the same in both
analyses, but a more detailed comparison is still missing.  Vice versa,
one should study if the constraints from supersymmetry, namely the
quaternionic geometry, are restrictive enough to fix the subleading
corrections that we have ignored. Experience from three-dimensional
gauge theories on the Coulomb branch suggests that we need some
additional information on the regularity and isometry structure of
the moduli space.

To make further progress, one clearly needs to embed our calculation
into string theory. Perhaps this can shed light on calculating the
one-loop determinant in the instanton background, or give more insight
into the structure of the subleading terms. Furthermore, from string
theory we learn that we should also take into account membrane instanton
effects. Just like for fivebranes, membrane instantons have a
supergravity description, as was demonstrated in \cite{TV1,DDVTV}. One
can therefore repeat our program for these solutions.

There are several interesting generalizations and applications, of which
we mention two. First, one can study the case where more than one
hypermultiplet is present. This corresponds to more general Calabi-Yau
manifolds with $h_{1,2}\neq 0$. Then, the effective action can also be
obtained from a type IIB compactification, or in some cases from the
heterotic string on $K3\times T^2$. In the latter case, one could use
the duality with the heterotic string to determine the hypermultiplet
moduli space, since on the heterotic side there are no corrections from
target space instantons. From the IIB perspective, one could study the
consequences of the $\SL(2,\fieldR)$ symmetry and of nonperturbative
mirror symmetry between IIA and IIB\@. Second, as an application, we
would like to compute the instanton corrections to the scalar potentials
that are obtained after gauging the unbroken isometries. Perhaps they
can lead to new interesting vacua with a nonvanishing cosmological
constant. We leave this for further research, and hope to report on this
in the future.

\bigskip
\textbf{Acknowledgments} 
\medskip

We would like to thank Lilia Anguelova, Tim Hollowood, Ruben Minasian
and Martin Ro\v{c}ek for stimulating discussions. UT and SV would like
to thank the organizers of the Workshop on Gravity in Two Dimensions
and the Erwin Schr\"odinger International Institute for Mathematical
Physics (ESI), where part of this work was done. UT is supported by the
Austrian Science Fund FWF, project no.\ P15553-N08.

\appendix 

\section{Spinor conventions} 
\label{conventions}

In this appendix, we elaborate on our conventions and properties of the
Euclidean theory. The spinors we work with are the continuation of
Lorentzian Weyl spinors $\la$, $\bla$ which are related by complex
conjugation. In Euclidean space, the Lorentz group becomes
$\mathrm{Spin}(4)=\SU(2)\times\SU(2)$, and the two-component spinors
$\la_\alpha$ and $\bla^{\alpha'}$ belong to inequivalent representations
of the two $\SU(2)$ factors, not related by complex conjugation.
Similarly for the Euclidean supersymmetry transformation parameters,
which are labeled by two independent two-component spinors $\eps^i$ and
$\beps_i$. As a consequence, the Euclidean action is not real, but it is
holomorphic in the spinors $\la$ and $\bla$.

The $\si^\mu$ and $\bsi^\mu$ matrices have lower and upper indices
respectively for their matrix entries, and we follow the notation
and conventions of Wess and Bagger \cite{WB}, adapted to Euclidean
space,
 \begin{equation} \label{Wicksigma}
  \si^\mu = (\vec{\si},-\I)\ ,\qquad \bsi^\mu = (-\vec{\sigma},-\I)\ ,
 \end{equation}
consistent with the identification $\si^4=\I\si^0$. This implies the
properties
 \begin{equation}
  \si^\mu \bsi^\nu = -g^{\mu\nu} + 2 \si^{\mu\nu}\ ,\qquad \half
  \ep^{\mu\nu\rho\si} \si_{\rho\si} = \si^{\mu\nu}\ ,
 \end{equation}
where $\si^{\mu\nu}\equiv\half\si^{[\mu}\bsi^{\nu]}$, and $\ep^{1234}=1$
($=e^{-1}$ in the local case). The second equation is a proper
self-duality equation, and differs by a factor of $\I$ from the
Lorentzian case. We also have
 \begin{equation}
  \si^\mu \bsi^\nu \si^\rho = g^{\mu\rho} \si^\nu - g^{\nu\rho} \si^\mu
  - g^{\mu\nu} \si^\rho + \ep^{\mu\nu\rho\si}\si_\si\ .
 \end{equation}
Further properties that are used are
 \begin{equation}\label{eps-sigma}
  (\si^\mu)_{\alpha\beta'}\, (\si_\mu)_{\gamma\alpha'} = - 2 \ep_{\alpha
  \gamma}\, \ep_{\beta'\alpha'}\ ,\qquad (\bsi^\mu)^{\beta'\alpha}\,
  (\si_\mu)_{\gamma\alpha'} = -2 \delta^\alpha_{\gamma} \delta^{\beta'
  }_{\alpha'}\ ,
\end{equation}
from which one can compute
 \begin{equation}
  (\si^{\mu\nu} \ep)_{\alpha\beta}\, (\si_\nu)_{\gamma\alpha'} = -
  \ep^{}_{\gamma(\beta}\, \si^\mu_{\alpha)\alpha'}\ ,
 \end{equation}
which is the same as in Lorentz signature.

\section{Dualization and target spaces}
\label{cosets}

As explained in the introduction, the dualization of scalars into
tensors changes the geometry of the remaining scalars. For
supergravities with maximal supersymmetry, this was demonstrated in
\cite{CLLPST,CJLP}. Here, we briefly mention the results for the coset
spaces that appear in our model. We distinguish between the Lorentzian
and Euclidean theories, since the signs of the pseudoscalars change
after Wick rotation. For the universal hypermultiplet, this leads to the
following two chains, corresponding to the hypermultiplet, the tensor
multiplet, and the double-tensor multiplet respectively. In the
Lorentzian theory, the duality chain is
 \begin{equation}
  \frac{\SU(1,2)}{\U(2)} \longrightarrow \frac{\SO(1,3)}{\SO(3)} \cong
  \frac{\SL(2,\fieldC)}{\SU(2)} \longrightarrow \frac{\SL(2,\fieldR)}
  {\Oo(2)}\ .
 \end{equation}
Notice that the scalar manifold for the tensor mulitplet is just
Euclidean $\mathrm{AdS}_3$.

For the Euclidean action, after Wick rotating the scalars, the sigma
model metric is no longer positive definite, and we have the duality
chain
 \begin{equation}
  \frac{\SL(3,\fieldR)}{\SL(2,\fieldR)\times\SO(1,1)} \longrightarrow
  \frac{\SO(2,2)}{\SO(2,1)} \cong \SL(2,\fieldR) \longrightarrow
  \frac{\SL(2,\fieldR)}{\Oo(2)}\ .
 \end{equation}
The tensor multiplet scalars, the middle step of the chain, now
correspond to $\mathrm{AdS}_3$.

The geometry of these scalar manifolds must be consistent with the
constraints from supersymmetry. These constraints are different in the
Lorentzian and Euclidean signatures. For hypermultiplets with Euclidean
supersymmetry, the target space is no longer quaternion-K\"ahler, as
already follows from the example given above. The precise constraints on
the geometry of the hypermultiplet scalars has, to the best of our
knowledge, not been worked out. For the scalars living in $N=2$ vector
multiplets, this was recently done in \cite{CMMS}, where it was shown
that the usual (special) K\"ahler geometry is replaced by (special)
para-K\"ahler geometry. We expect that for hypermultiplets the geometry
of quaternion-K\"ahler manifolds will be replaced by the notion of
para-quaternionic K\"ahler geometry. For some mathematics literature on
this, see e.g.\ \cite{GRMVL}.

\section{Constraints in the scalar-tensor models} 
\label{sc-tens-rel}

We here present the constraints on and relations between the various
quantities appearing in the action \eqref{Eucl-rigid-action} and
supersymmetry transformation rules \eqref{susy-dtm} of the scalar-tensor
system. These all follow from the requirement of the closure of the
supersymmetry algebra and the invariance of the action, and were
discussed extensively in \cite{TV2}.

The algebraic relations one finds are
\begin{gather}
  \ga_{ia}^A\, \W{A}{bj} + g_{Iia}\, f^{Ibj} = \de_i^j\, \de_a^b
    \notag \\[2pt]
  \ga_{ia}^A\, \bW{Aj}{\ab} + g_{Iia}\, \bar{f}^{I\ab}{}_j + (i
    \leftrightarrow j) = 0\ , \label{rel1}
 \end{gather}
for contractions over $A$ and $I$, and
 \begin{equation}
  \begin{pmatrix} \ga_{ia}^A\, \W{B}{aj} & \ga_{ia}^A\, f^{Jaj} \\[2pt]
  g_{Iia}\, \W{B}{aj} & g_{Iia}\, f^{Jaj} \end{pmatrix} + \text{c.c.}
  (i \leftrightarrow j) = \de_i^j \begin{pmatrix} \de_B^A & 0 \\[2pt]
  0 & \de_I^J \end{pmatrix}\ . \label{rel2}
 \end{equation}
Furthemore, we have
\begin{equation} \label{Gga_Mg}
  \cG_{AB}\, \ga_{ia}^B = h_{a\bb}\, \bW{Ai}{\bb}\ ,\qquad M^{IJ}
  g_{Jia} = h_{a\ab} \bar{f}^{I\ab}{}_i\ .
 \end{equation}
These relations imply among others that
 \begin{gather}
  \cG_{AB} = h_{a\bb}\, \W{A}{ai}\, \bW{Bi}{\bb}\ ,\qquad
    M^{IJ} = h_{a\bb}\, f^{Iai} \bar{f}^{J\bb}{}_i \notag \\[2pt]
  \de_i^j\, h_{a\bb} = \cG_{AB}\, \ga^A_{ia}\, \bar{\ga}^j{}^B_\bb
    + M^{IJ} g_{Iia}\, \bar{g}^j_{J\bb} \ . \label{GMh}
 \end{gather}
For the complete set of relations, involving the covariant derivatives
and curvatures, we refer to \cite{TV2}. We mention here only some
relations useful for later purposes. Of particular interest is the
following:
 \begin{equation} \label{F}
  F_{AB}{}^I = 2\I M^{IJ} k_{Ja\ab} \bW{Ai}{\ab}\, \W{B}{ai} - 2
  h_{a\ab} \bW{Ai}{\ab}\, \Gamma^{Ii}{}_j \W{B}{aj}\ ,
 \end{equation}
where the field strength is defined as ${F_{AB}}^I=2\partial^{}_{[A}
A_{B]}^I$. This field strength also measures the nonvanishing of the
covariant derivatives of $\gamma^A_{ia}$ and $\W{A}{ai}$. The higher
order fermion terms in the supersymmetry transformation rules
\eqref{susy-dtm} contain tensors $\Gamma^{Ia}{}_b$ that satisfy
 \begin{equation} \label{k-Gamma}
  M^{IJ} k_{Ja\ab} = \I h_{b\ab} \Gamma^{Ib}{}_a\ .
 \end{equation}

Finally, one can define a covariantly constant tensor
\begin{equation} \label{def-E-tensor}
  \cE_{ab} = \half \ep^{ji}\, (\cG_{AB}\, \ga^A_{ia}\, \ga^B_{jb}\
  + M^{IJ} g_{Iia}\, g_{Jjb})\ ,
 \end{equation}
which satisfies
 \begin{equation} \label{EWidentity}
  \cE_{ab} \lp \W{A}{bi} \\[2pt] f^{Ibi} \rp = \ep^{ij} h_{a\ab}
  \lp \bW{Aj}{\ab} \\[2pt] \bar{f}^{I\ab}{}_j \rp\ .
 \end{equation}
This tensor appears explicitly in the four-fermi terms in the
supergravity action \eqref{EuclDTMaction}.

\section{Coefficients for the double-tensor multiplet} 
\label{coeff_DTM}

The universal double-tensor multiplet provides a solution to the
constraints in the previous section. In the following we list the
coefficient functions appearing in its classical action and
transformation laws which satisfy the relations given above. They
receive quantum corrections from instantons, some determined in this
paper, and 1-loop effects \cite{AMTV}.

For the scalar zweibeins we have
 \begin{equation}
  \gamma^\phi_{ia} = (W_\phi^{ai})^\dag = \frac{1}{\2} \begin{pmatrix}
  0 & -1 \\[2pt] 1 & 0 \end{pmatrix}\ ,\qquad \gamma^\chi_{ia} = \e^\phi
  (W_\chi^{ai})^\dag = \frac{1}{\2}\, \e^{\phi/2} \begin{pmatrix} 1 & 0
  \\[2pt] 0 & 1 \end{pmatrix}\ ,
 \end{equation}
while the tensor zweibeins are given by, for $I=1,2$,
 \begin{equation}
  g_{1\,ia} = -\frac{\I}{\2}\, \e^{-\phi} \begin{pmatrix} -\e^{\phi/2} &
  \chi \\[2pt] \chi & \e^{\phi/2} \end{pmatrix}\ ,\qquad g_{2\,ia} = -
  \frac{\I}{\2}\, \e^{-\phi} \begin{pmatrix} 0 & 1 \\[2pt] 1 & 0
  \end{pmatrix}\ ,
 \end{equation}
and
 \begin{equation}
  f^{1\,ai} = \frac{\I}{\2}\, \e^{\phi/2} \begin{pmatrix} -1 & 0 \\[2pt]
  0 & 1 \end{pmatrix}\ ,\qquad f^{2\,ai} = \frac{\I}{\2}\, \e^{\phi/2}
  \begin{pmatrix} \chi & \e^{\phi/2} \\[2pt] \e^{\phi/2} & -\chi
  \end{pmatrix}\ .
 \end{equation}
One may check that these quantities satisfy the relations
\eqref{rel1}--\eqref{EWidentity}, with
 \begin{equation}
  h_{a\ab}=\begin{pmatrix} 1 & 0 \\[2pt] 0 & 1 \end{pmatrix}\ ,\qquad
  \cE_{ab}=\begin{pmatrix} 0 & -1 \\[2pt] 1 & 0 \end{pmatrix}\ ,
 \end{equation}
where we also have taken $\ep^{12}=1$.

The target space connections for the double-tensor multiplet are
particularly simple:
 \begin{equation}\label{DTM-connections}
  \G{A}{a}{b} = 0\ ,\qquad \G{\phi}{i}{j} = 0\ ,\qquad \G{\chi}{i}{j}
  = \frac{1}{2}\, \e^{-\phi/2} \lp 0 & -1 \\[2pt] 1 & 0 \rp\ .
 \end{equation}
Since $F_{AB}{}^I=0$, the scalar zweibeins $\W{A}{ai}$, $\ga^A_{ia}$
are covariantly constant with respect to these connections. The tensor
$k_{Ia\ab}$ can be determined from \eqref{k-Gamma}, with
 \begin{equation} \label{Gamma2ab}
  \Gamma^{1\,a}{}_b = 0\ ,\qquad \Gamma^{2\,a}{}_b = -\frac{3\I}{4}\,
  \e^\phi \lp 1 & 0 \\[2pt] 0 & -1 \rp\ .
 \end{equation}
Other quantities are the gravitino coefficients in the supersymmetry
transformations of the tensors \eqref{de_B_loc}
 \begin{equation}
  {\Omega_1}^i{}_j = \frac{\I}{4}\, \e^{-\phi} \lp \chi & 2\,
  \e^{\phi/2} \\[2pt] 2\, \e^{\phi/2} & -\chi \rp\ ,\qquad
  {\Omega_2}^i{}_j = \frac{\I}{4}\, \e^{-\phi} \lp 1 & 0 \\[2pt] 0 &
  -1 \rp\ ,
 \end{equation}
and, by using \eqref{Gamma-Omega}, the coefficients in the
transformations of the gravitinos
 \begin{equation}
  \Gamma^{1\,i}{}_j = \frac{\I}{2}\, \e^{\phi/2} \begin{pmatrix} 0 & 1
  \\[2pt] 1 & 0 \end{pmatrix}\ ,\qquad \Gamma^{2\,i}{}_j = - \frac{\I}
  {4}\, \e^{\phi/2} \begin{pmatrix} -\e^{\phi/2} & 2 \chi \\[2pt] 2 \chi
  & \e^{\phi/2} \end{pmatrix}\ .
 \end{equation}
Just as in the universal hypermultiplet, the four-$\la$ terms come with
field-independent coefficients,
 \begin{equation} \label{cR_DTM}
  \frac{1}{4}\, V_{ab\,\ab\bb}\, \la^a \la^b\, \bla^\ab \bla^\bb = -
  \frac{3}{8}\, \big( \la^1 \la^1\, \bla^1 \bla^1 - 2\, \la^1 \la^2\,
  \bla^1 \bla^2 + \la^2 \la^2\, \bla^2 \bla^2 \big)\ .
 \end{equation}

\section{Zero mode norms} 
\label{formulas}

In the computation of zero mode norms we frequently encounter integrals
of the form
 \begin{equation*}
  I_p = \int\! \d^4x\, h^{-p}\, (\p_r h)^2\ ,\qquad h = h_\infty +
  \frac{Q}{4\pi^2\,r^2}\ ,
 \end{equation*}
for some power $p>1$. Using $\d r\,r^{-3}=-2\pi^2\d h/Q$, this turns
into
 \begin{equation} \label{I_p}
  I_p = \Big(\frac{Q}{2\pi^2}\Big)^{\!2}\, \mathrm{Vol}(S^3)
  \int_0^\infty\!\! \d r\, r^{-3}\, h^{-p} = Q \int_{h_\infty}^\infty
  \!\! \d h\, h^{-p} = \frac{Q}{p-1}\, h_\infty^{1-p}\ .
 \end{equation}
The integral diverges for $p\leq 1$.

\section{Instanton corrections to the vielbeins} 
\label{inst-vielb}

Although we are mostly interested in instanton corrections to the scalar
and tensor metrics, it is worthwhile also to compute the corrected
zweibeins $\W{A}{ai}$ etc. Once these are known, one can compute the
vierbeins on the quaternionic side by using the results of \cite{TV2}.
Obviously, they are determined only up to $\SU(2)$ rotations. We use
this fact to choose the components as simple as possible.

We begin with determining $\W{A}{ai}$ from the first relation in
\eqref{GMh}, $\cG_{AB}=\Tr(W_{\!A}^\dag\,h^t\,W_{\!B}^{})$. Since as
noted above both $\cE_{ab}$ and $h_{a\ab}$ are given by their classical
expressions, \eqref{EWidentity} implies that $W_{\!A}$ (and $f^I$)
are of the form
 \begin{equation*}
  \W{A}{ai} = \lp u_A & v_A \\[2pt] - \bar{v}_A & \bar{u}_A \rp\ .
 \end{equation*}
We solve $\cG_{\phi\phi}^\inst=\cG_{\phi\chi}^\inst=0$ by setting
$W_{\!\phi}^\inst=0$. Furthermore, we can choose $u_\chi$ real and
$v_\chi=0$; then $\cG_{\chi\chi}^\inst=-g_s^2 Y$ implies
 \begin{equation} 
  W_{\!\chi}^{\eff\,ai} = \frac{1}{2\2}\, \big( 2 \e^{-\phi/2} - g_s
  Y \big) \lp 1 & 0 \\[2pt] 0 & 1 \rp\ .
 \end{equation}
The $\ga^A$ now follow from \eqref{Gga_Mg}, $\ga^A=\cG^{AB}W_{\!B}^\dag
\,h^t$,
 \begin{equation}
  \ga^\phi_{\eff\,ia} = \ga^\phi_{ia}\ ,\qquad \ga^\chi_{\eff\,ia}
  = \frac{1}{2\2}\, \big( 2 \e^{\phi/2} + g_s^{-1} Y \big) \lp 1 & 0
  \\[2pt] 0 & 1 \rp\ .
 \end{equation}

The determination of $g_I$ is analogous to the one of $W_{\!A}$;
from $M_{IJ}=\Tr(g_I^\dag\,g_J^{}\,h^{-1t})$ and $M^\inst_{12}=
M^\inst_{22}=0$ we conclude that $g^\inst_2=0$. For $g_1$ we take
 \begin{equation}
  g_{1ia}^\eff = g_{1ia} + \frac{\I}{2\2}\, g_s Y \lp 1 & 0 \\[2pt] 0
  & -1 \rp\ .
 \end{equation}
The coefficients $f^I$ finally are given by the relation $f^I=M^{IJ}
g_J^\dag\,h^{-1t}$:
 \begin{equation}
  f^{1ai}_\eff = f^{1ai} + \frac{\I}{2\2}\, g_s^{-1} Y \lp 1 & 0
  \\[2pt] 0 & -1 \rp\ ,\quad f^{2\,ai}_\eff = f^{2\,ai} - \frac{\I}
  {2\2}\, \chi_\infty\, g_s^{-1} Y \lp 1 & 0 \\[2pt] 0 & -1 \rp\ .
 \end{equation}

In section \ref{sect_modul} we have observed that the NS sector in the
effective action is not affected by instantons. With the above vielbeins
we can check whether this might hold for the supersymmetry
transformations as well. Indeed, $g_2^\inst=0$ and $\Omega_2^\inst=0$
imply that the transformation \eqref{de_B_loc} of the NS 2-form
$B_{\mu\nu2}$, dual to the axion $\si$, is not corrected, while
$\ga^\phi_\inst=0$ leaves the dilaton transformation unchanged.

\raggedright


\begin{thebibliography}{99} 

\bibitem{S-CHS1-DL}
A.~Strominger, \emph{Heterotic solitons.} Nucl.\ Phys.\ \textbf{B343}
(1990) 167. [Erratum: \textbf{B353} (1991) 565]; \\
C.~Callan, Jr., J.~Harvey and A.~Strominger, \emph{World sheet approach
to heterotic instantons and solitons.} Nucl.\ Phys.\ \textbf{B359}
(1991) 611; \\
M.~Duff and J.~Lu, \emph{Elementary five-brane solutions of D=10
supergravity.} Nucl.\ Phys.\ \textbf{B354} (1991) 141.

\bibitem{CHS2}
C.~Callan, Jr., J.~Harvey and A.~Strominger, \emph{Worldbrane actions
for string solitons.} Nucl.\ Phys.\ \textbf{B367} (1991) 60.

\bibitem{DVV}
R.~Dijkgraaf, E.~Verlinde and M.~Vonk, \emph{On the partition sum of
the NS five-brane.} \arxiv{hep-th/0205281}.

\bibitem{HM-GKKOPP}
J.~Harvey and G.~Moore, \emph{Fivebrane instantons and $R^2$ couplings
in $N=4$ string theory.} Phys.\ Rev.\ \textbf{D57} (1998) 2323,
\arxiv{hep-th/9610237}; \\
A.~Gregori, E.~Kiritsis, C.~Kounnas, N.~Obers, P.~Petropoulos and
B.~Pioline, \emph{$R^2$ corrections and nonperturbative dualities
of $N=4$ string ground states.} Nucl. Phys. \textbf{B510} (1998) 423,
\arxiv{hep-th/9708062}.

\bibitem{KOP}
E.~Kiritsis, N.~Obers and B.~Pioline, \emph{Heterotic/Type II triality
and instantons on $K3$.} JHEP \textbf{0001} (2000) 029,
\arxiv{hep-th/0001083}.

\bibitem{Ki}
E.~Kiritsis, \emph{Duality and instantons in string theory.} In
``Trieste 1999, Superstrings and related matters'', 127-205.
\arxiv{hep-th/9906018}.

\bibitem{BBS}
K.~Becker, M.~Becker and A.~Strominger, \emph{Fivebranes, membranes
and nonperturbative string theory.} Nucl.\ Phys.\ \textbf{B456}
(1995) 130, \arxiv{hep-th/9507158}.

\bibitem{R}
S.-J.~Rey, \emph{The confining phase of superstrings and axionic
strings.} Phys.\ Rev.\ \textbf{D43} (1991) 526; \\
S.-J.~Rey and T.R.~Taylor, \emph{Instanton effects in supergravity
theories.} Phys.\ Rev.\ Lett.\ \textbf{71} (1993) 1132,
\arxiv{hep-th/9305064}.

\bibitem{W}
E.~Witten, \emph{Non-perturbative superpotentials in string theory.}
Nucl.\ Phys.\ \textbf{B474} (1996) 343, \arxiv{hep-th/9604030}.

\bibitem{BCF}
S.~Ferrara and S.~Sabharwal, \emph{Dimensional reduction of type~II
superstrings.} Class.\ Quantum.\ Grav.\ \textbf{6} (1989) L77; \\
M.~Bodner, A.~Cadavid and S.~Ferrara, \emph{(2,2) Vacuum configurations
for type~IIA superstrings: $N=2$ supergravity Lagrangians and algebraic
geometry.} Class.\ Quantum.\ Grav.\ \textbf{8} (1991) 789.

\bibitem{A}
P.~Aspinwall, \emph{Compactification, geometry and duality: N=2.}
In ``Boulder 1999, Strings, branes and gravity'', 723.
\arxiv{hep-th/0001001}.

\bibitem{BB}
K.~Becker and M.~Becker, \emph{Instanton action for type II
hypermultiplets.} Nucl.\ Phys.\ \textbf{B551} (1999) 102,
\arxiv{hep-th/9901126}.

\bibitem{CFG}
S.~Cecotti, S.~Ferrara and L.~Girardello, \emph{Geometry of type~II
superstrings and the moduli of superconformal field theories.} Int.\
J.\ Mod.\ Phys.\ \textbf{A4} (1989) 2475; \\
S.~Ferrara and S.~Sabharwal, \emph{Quaternionic manifolds for type~II
superstring vacua of Calabi-Yau spaces.} Nucl.\ Phys.\ \textbf{B332}
(1990) 317.

\bibitem{BW}
J.~Bagger and E.~Witten, \emph{Matter couplings in $N=2$ supergravity.}
Nucl.\ Phys.\ \textbf{B222} (1983) 1.

\bibitem{AMTV}
I.~Antoniadis, R.~Minasian, S.~Theisen and P.~Vanhove, \emph{String loop
corrections to the universal hypermultiplet.} Class.\ Quant.\ Grav.\
\textbf{20} (2003) 5079, \arxiv{hep-th/0307268}.

\bibitem{S}
A.~Strominger, \emph{Loop corrections to the universal hypermultiplet.}
Phys.\ Lett.\ \textbf{B421} (1998) 139, \arxiv{hep-th/9706195}; \\
I.~Antoniadis, S.~Ferrara, R.~Minasian and K.S.~Narain, \emph{$R^4$
couplings in M and type~II theories.} Nucl.\ Phys.\ \textbf{B507}
(1997) 571, \arxiv{hep-th/9707013}.

\bibitem{GHL}
H.~G\"unther, C.~Herrmann and J.~Louis, \emph{Quantum corrections
in the hypermultiplet moduli space.} Fortsch. Phys. \textbf{48} (2000)
119, \arxiv{hep-th/9901137}.

\bibitem{ARV}
L.~Anguelova, M.~Ro\v{c}ek and S.~Vandoren, \emph{Quantum corrections to
the universal hypermultiplet and superspace.} \arxiv{hep-th/0402132}.

\bibitem{dWRV}
B.~de~Wit, M.~Ro\v{c}ek and S.~Vandoren, \emph{Hypermultiplets,
hyperk\"ahler cones and quaternionic-k\"ahler geometry.} JHEP
\textbf{0102} (2001) 039, \arxiv{hep-th/0101161}.

\bibitem{GMV}
B.R.~Greene, D.R.~Morrison, C.~Vafa, \emph{A geometric realization of
confinement.} Nucl.\ Phys.\ \textbf{B481} (1996) 513,
\arxiv{hep-th/9608039}.

\bibitem{OV}
H.~Ooguri and C.~Vafa, \emph{Summing up D-instantons.} Phys.\ Rev.\
Lett.\ \textbf{77} (1996) 3296, \arxiv{hep-th/9608079}.

\bibitem{GS}
M.~Gutperle and M.~Spalinski, \emph{Supergravity instantons and the
universal hypermultiplet.} JHEP \textbf{0006} (2000) 037,
\arxiv{hep-th/0005068}; \emph{Supergravity instantons for $N=2$
hypermultiplets.} Nucl.\ Phys.\ \textbf{B598} (2001) 509,
\arxiv{hep-th/0010192}.

\bibitem{K}
S.~V.~Ketov, \emph{Universal hypermultiplet metrics.} Nucl.\ Phys.\
\textbf{B604} (2001) 256, \arxiv{hep-th/0102099}; \emph{Summing up
D-instantons in N=2 supergravity.} Nucl.\ Phys.\ \textbf{B649} (2003)
365, \arxiv{hep-th/0209003}; \emph{D-instantons and matter
hypermultiplet.} Phys.\ Lett.\ \textbf{B558} (2003) 119,
\arxiv{hep-th/0302001}.

\bibitem{SW}
N.~Seiberg and E.~Witten, \emph{Gauge dynamics and compactification
to three dimensions.} In \emph{The mathematical beauty of physics},
p.~333, Eds.\ J.M.~Drouffe and J.-B.~Zuber (World Scient., 1997),
\arxiv{hep-th/9607163}; \\
G.~Chalmers and A.~Hanany, \emph{Three-dimensional gauge theories and
monopoles.} Nucl.\ Phys.\ \textbf{B489} (1997) 223,
\arxiv{hep-th/9611063}.

\bibitem{DKMTV}
N.~Dorey, V.V.~Khoze, M.P.~Mattis, D.~Tong and  S.~Vandoren,
\emph{Instantons, three-dimensional gauge theory, and the
Atiyah-Hitchin manifold.} Nucl.\ Phys.\ \textbf{B514} (1998) 553,
\arxiv{hep-th/9703228}; \\
C.~Fraser and D.~Tong, \emph{Instantons, three-dimensional gauge
theories, and monopole moduli spaces.} Phys.\ Rev.\ \textbf{D58}
(1998) 085001, \arxiv{hep-th/9710098}; \\
N.~Dorey, D.~Tong and S.~Vandoren, \emph{Instanton effects in
three-dimensional supersymmetric gauge theories with matter.}
JHEP \textbf{9804} (1998) 005, \arxiv{hep-th/9803065}.

\bibitem{BM}
K.~Behrndt and S.~Mahapatra, \emph{De Sitter vacua from $N=2$ gauged
supergravity.} JHEP \textbf{0401} (2004) 068, \arxiv{hep-th/0312063}.

\bibitem{TV1}
U.~Theis and S.~Vandoren, \emph{Instantons in the double-tensor
multiplet.} JHEP \textbf{0209} (2002) 059, \arxiv{hep-th/0208145}.

\bibitem{DDVTV}
M.~Davidse, M.~de Vroome, U.~Theis and S.~Vandoren, \emph{Instanton
solutions for the universal hypermultiplet.} Fortschr.\ Phys.\
\textbf{52} (2004) 696, \arxiv{hep-th/0309220}.

\bibitem{TV2}
U.~Theis and S.~Vandoren, \emph{N=2 supersymmetric scalar-tensor
couplings.} JHEP \textbf{0304} (2003) 042, \arxiv{hep-th/0303048}.

\bibitem{GGP}
G.W.~Gibbons, M.B.~Green and M.J.~Perry, \emph{Instantons and
seven-branes in type II-B superstring theory.} Phys.\ Lett.\
\textbf{B370} (1996) 37, \arxiv{hep-th/9701093}; \\
M.B.~Green and M.~Gutperle, \emph{Effects of D-instantons.} Nucl.\
Phys.\ \textbf{B498} (1997) 195, \arxiv{hep-th/9701093}.

\bibitem{CLLPST}
E.~Cremmer, I.V.~Lavrinenko, H.~L\"u, C.N.~Pope, K.S.~Stelle and
T.A.~Tran, \emph{Euclidean-signature supergravities, dualities and
instantons.} Nucl.\ Phys.\ \textbf{B534} (1998) 40,
\arxiv{hep-th/9803259}.

\bibitem{P}
A.~Polyakov, \emph{Quark confinement and topology of gauge theories.}
Nucl.\ Phys.\ \textbf{B120} (1977) 429; see also chapter 4 in
\emph{Gauge fields and strings.} Harwood academic publishers of
``Contemporary concepts in physics'', Volume 3, 1987.

\bibitem{TvN}
U.~Theis and P.~van~Nieuwenhuizen, \emph{Ward identities for $N=2$
rigid and local supersymmetry in Euclidean space.} Class.\ Quant.\
Grav.\ \textbf{18} (2001) 5469, \arxiv{hep-th/0108204}.

\bibitem{B}
C.~Bernard, \emph{Gauge zero modes, instanton determinants, and QCD
calculations.} Phys.\ Rev.\ \textbf{D19} (1979) 3013.

\bibitem{BVvN}
A.~Belitsky, S.~Vandoren and P.~van~Nieuwenhuizen, \emph{Yang-Mills-
and D-instantons.} Class.\ Quant.\ Grav.\ \textbf{17} (2000) 3521,
\arxiv{hep-th/0004186}.

\bibitem{HM2}
J.~Harvey and G.~Moore, \emph{Superpotentials and membrane instantons.}
\arxiv{hep-th/9907026}.

\bibitem{DHKM}
N.~Dorey, V.~V.~Khoze and M.~P.~Mattis, \emph{Multi-Instanton calculus
in $N=2$ supersymmetric gauge theory.} Phys.\ Rev.\ \textbf{D54} (1996)
2921, \arxiv{hep-th/9603136}; \\
N.~Dorey, T.~J.~Hollowood, V.~V.~Khoze, M.~P.~Mattis and S.~Vandoren,
\emph{Multi-Instanton calculus and the AdS/CFT correspondence in N=4
superconformal field theory.} Nucl.\ Phys.\  \textbf{B552} (1999) 88,
\arxiv{hep-th/9901128}; \\
N.~Dorey, T.~J.~Hollowood, V.~V.~Khoze and M.~P.~Mattis, \emph{The
calculus of many instantons.} Phys.\ Rept.\ \textbf{371} (2002) 231,
\arxiv{hep-th/0206063}.

\bibitem{DWVP}
B.~de Wit and A.~Van Proeyen, \emph{Symmetries of dual-quaternionic
manifolds.} Phys.\ Lett.\ \textbf{B252} (1990) 221.

\bibitem{DWVVP}
B.~de Wit, F.~Vanderseypen and A.~Van Proeyen, \emph{Symmetry structure
of special geometries.} Nucl.\ Phys.\ \textbf{B400} (1993) 463,
\arxiv{hep-th/9210068}.

\bibitem{WB}
J.~Wess and J.~Bagger, \emph{Supersymmetry and Supergravity.} Princeton
Series in Physics (Princeton Univ.\ Press, Princeton, 1983).

\bibitem{CJLP}
E.~Cremmer, B.~Julia, H.~L\"u and C.N.~Pope, \emph{Dualisation of
dualities. I.} Nucl.\ Phys.\ \textbf{B523} (1998) 73,
\arxiv{hep-th/9710119}.

\bibitem{CMMS}
V.~Cort\'es, C.~Mayer, T.~Mohaupt and F.~Saueressig, \emph{Special
geometry of euclidean supersymmetry I: Vector multiplets.} JHEP
\textbf{0403} (2004) 028, \arxiv{hep-th/0312001}.

\bibitem{GRMVL}
E.~Garcia-Rio, Y.~Matsushita, R.~Vasquez-Lorenzo, \emph{Paraquaternionic
K\"ahler manifolds.} Rocky Mountain J. Math. \textbf{31} (2001) 237; \\
S.~Vukmirovi\'c, \emph{Para-quaternionic reduction.}
\arxiv{math.DG/0304424}.

\end{thebibliography}
\end{document}